\documentclass[10pt,twocolumn]{IEEEtran}
\usepackage{amssymb}
\usepackage{amsmath}
\usepackage[version=3]{mhchem}
\usepackage{graphicx}
\usepackage{multirow}
\usepackage{wrapfig}
\usepackage{color}
\usepackage{colortbl}
\usepackage{soul}
\usepackage{subfigure}
\usepackage{enumerate}
\usepackage{paralist}
\usepackage[normalem]{ulem}
\usepackage{mathtools}

\DeclareMathOperator*{\argmax}{arg\,max}
\DeclareMathOperator\erfc{erfc}

\DeclarePairedDelimiter{\ceil}{\lceil}{\rceil}

\begin{document}
\title{Modeling and Analysis of SiNW FET-Based\\Molecular Communication Receiver}
\author{Murat Kuscu,~\IEEEmembership{Student Member,~IEEE}
        and Ozgur B. Akan,~\IEEEmembership{Fellow,~IEEE}
        \thanks{An earlier version of this work \cite{Kuscu2015c} was presented at IEEE WF-IoT'15, Milan, Italy.}
        \thanks{The authors are with the Next-generation and Wireless Communications Laboratory (NWCL), Department of Electrical and Electronics Engineering, Koc University, Istanbul, 34450, Turkey (e-mail: \{mkuscu, akan\}@ku.edu.tr).}
        \thanks{This work was supported in part by the ERC project MINERVA (ERC-2013-CoG \#616922), and the EU project CIRCLE (EU-H2020-FET-Open \#665564).}}

\maketitle

\begin{abstract}
Molecular Communication (MC) is a bio-inspired communication method based on the exchange of molecules for information transfer among nanoscale devices. MC has been extensively studied from various aspects in the literature; however, the physical design of MC transceiving units is largely neglected with the assumption that network nodes are entirely biological devices, e.g., engineered bacteria, which are intrinsically capable of receiving and transmitting molecular messages. However, the low information processing capacity of biological devices and the challenge to interface them with macroscale networks hinder the true application potential of nanonetworks. To overcome this limitation, recently, we proposed a nanobioelectronic MC receiver architecture exploiting the nanoscale field effect transistor-based biosensor (bioFET) technology, which provides noninvasive and sensitive molecular detection while producing electrical signals as the output. In this paper, we introduce a comprehensive model for silicon nanowire (SiNW) FET-based MC receivers by integrating the underlying processes in MC and bioFET to provide a unified analysis framework. We derive closed-form expressions for the noise statistics, the signal-to-noise ratio (SNR) at the receiver output, and the symbol error probability (SEP). Performance evaluation in terms of SNR and SEP reveals the effects of individual system parameters on the detection performance of the proposed MC receiver.
\end{abstract}

\begin{IEEEkeywords}
Molecular communication, receiver, SiNW, bioFET, SNR, SEP.
\end{IEEEkeywords}

\IEEEpeerreviewmaketitle

\section{Introduction}
\label{Introduction}
\IEEEPARstart{M}{olecular} communication (MC) defines the technology where molecules are used to encode, transmit, and receive information. It is a biocompatible method providing efficient and reliable information transfer between living entities at nanoscale. Hence, it is regarded as the most promising paradigm to enable nanonetworks, i.e., network of nanoscale devices \cite{Akan2016}-\cite{Kuscu2015n}.

MC has been extensively studied from various aspects. A large body of work has been devoted to modeling the MC channel from information theoretical perspective \cite{Akyildiz2013}, designing modulation schemes \cite{Nakano2012} and developing communication protocols \cite{Nakano2014} and optimal detection algorithms \cite{Kilinc2013}. When adapting the tools of conventional communication techniques to MC, these studies have mostly ignored the physical design of system components such as transmitter and receiver. There are only a few studies focusing on the physical aspects of communication units \cite{Nakano2014}, \cite{Unluturk2015}, \cite{Okonkwo2015}. For example, in \cite{Unluturk2015}, the authors propose a biotransceiver architecture which can realize transmitting, receiving and basic processing operations based on the functionalities of genetically engineered bacteria. Similarly, a layered architecture for MC based on functionalities of bio-nanomachines, e.g., molecular motors, is presented in \cite{Nakano2014}. Furthermore, numerous studies have investigated MC-based networks of bacteria colonies \cite{Sasi2015}. Common to these studies is the assumption that the MC nanonetwork consists of nanomachines which are entirely made up of biological components.

Although designing nanomachines with only bio-components provides the advantage of biocompatibility, it has also numerous disadvantages restricting the application domain of nanonetworks. First of all, very low computational capacities of bio-devices limit the speed of information processing, and thus, the extent of the tasks that can be accomplished \cite{Unluturk2015}. This limitation points out a major discrepancy considering the envisaged applications of nanonetworks \cite{Akan2016}, most of which require the implementation of complex communication protocols and algorithms. Another critical drawback of entirely biological architectures is that they are only operational in \emph{in vivo} applications, i.e., applications within living organisms. Moreover, they do not allow the incorporation of a noninvasive and seamless interface between molecular nanonetworks and macroscale cyber networks such as the Internet. This is one of the key challenges to realization of the Internet of nanothings (IoNT) \cite{Akyildiz2015}, \cite{Kuscu2015a}.

The discrepancies and challenges pertaining to entirely biological architectures have led us to consider different solutions. In our recent review of design options \cite{Kuscu2015b}, interfacing the biochemical environment, where molecular messages propagate, with a nanobioelectronic architecture providing fast information processing and wireless macro-nano interface has been revealed to be the most promising and feasible solution. Nanobioelectronic design implies transmitters that can release molecular messages upon being triggered by electrical signals, and receivers that can detect molecular messages and transduce them into electrical signals for further processing.

In \cite{Kuscu2015b}, for implementing a nanobioelectronic MC receiver, we proposed the use of FET-based biosensors, i.e., bioFETs, optimized from MC theoretical perspective. BioFETs have emerged as promising analytical tools, which enable the label-free electrical sensing of target molecules \cite{Poghossian2014}. They satisfy the basic requirements of an MC receiver, such as the capability of precise, continuous and noninvasive detection of molecular concentrations. Use of novel nanomaterials, such as nanowires, carbon nanotube (CNT) and graphene, has enabled the design of bioFETs with nanoscale dimensions \cite{Curreli2008}. Transduction of biochemical concentrations into electrical signals in bioFETs could provide a fast in-device information processing for the MC receiver. It could also enable the design of a seamless interface between the nanobioelectronic receiver and macroscale networks by means of electromagnetic communication. Towards the goal of optimizing bioFETs as MC receivers, one of the major challenges pointed out in \cite{Kuscu2015b} is the lack of a comprehensive analytical model for bioFETs. Although there are a vast number of experimental works reported for bioFETs and a few theoretical studies focusing on the noise processes effective on their operation \cite{Deen2006}, \cite{Rajan2013}, none of them is able to entirely capture the physical processes in stochastic sensing of molecular concentration.

In \cite{Kuscu2015b} and \cite{Kuscu2015c}, we introduced a modeling approach for studying the performance of SiNW FET-based MC receiver antennas, and derived output SNR and receiver sensitivity metrics based on a number of simplifying assumptions and considering a free diffusion channel. Unfortunately the presented preliminary model lacks the ability to reflect the correlations between critical system parameters and capture the geometrical peculiarities of SiNWs; and thus, could only provide a very limited applicability. In this study, we consider a microfluidic advection-diffusion channel for the propagation of information ligands, and revisit and largely extend the receiver model especially by accounting for the correlations between the ligand transport dynamics and the ligand-receptor binding kinetics. The model also captures the correlations among the transducer capacitances, geometrical properties and operating voltages of SiNW FET, and the controllable system parameters, such as the ionic strength of the propagation channel and temperature. The presented model provides analytical expressions for the received signal and noise statistics at the receiver's electrical output as a function of a large set of independent system parameters. Based on the received signal statistics, we derive an analytical expression for the symbol error probability (SEP) considering an MC system that utilizes M-ary concentration shift keying (M-CSK) \cite{Nakano2012} for modulation and maximum likelihood (ML) method for detection. This contribution ultimately enables an analysis and optimization framework revealing the impact of individual parameters on the communication performance. The performance evaluation results obtained based on the introduced model already outline the feasible optimization pathways that can be targeted to improve the receiver.

In the presented model, we incorporate the spatial and temporal correlation effects resulting from finite-rate transport of ligands to the stochastic ligand-receptor binding process, and obtain the statistics of the transport-influenced binding noise at the receiver. This development leads to a combined channel and receiver noise model, which can be applicable for MC systems with various receiver architectures, including the entirely biological ones. Up to now, a vast majority of MC studies focus on the diffusion-limited regime by assuming that the receiver could perfectly count the number of molecules in an arbitrarily defined reception space \cite{Akyildiz2013}, \cite{Noel2014}. There are also a few studies, such as \cite{Pierobon2011}, that focus on the reaction-limited case and derive the statistics of ligand-receptor binding process, which is Markovian in this regime. However, both approaches neglect the extended correlations in the reception process resulting from the finite transport and reaction rates. There is another modeling approach based on reaction-diffusion master equation (RDME), which provides a combined channel-receiver model by taking the propagation and binding process as a continuous-time Markov process (CTMP) \cite{Chou2013}, \cite{Chou2015}. The developed model is comprehensive in its approach to small-scale systems; however, it does not allow the derivation of closed-form expressions for noise statistics. The unified noise model developed in this paper overcomes this limitation with a steady-state assumption in the received concentration signal.

The use of electronic sensors for MC receiver has been previously addressed in \cite{Farsad2013}. In that study, the authors designed a testbed for MC with macroscale dimensions and use a metal oxide semiconductor sensor for detecting chemical messages encoded into isopropyl alcohol. Based on the experimental data obtained using the testbed, they developed a combined channel and receiver model \cite{Kim2015}. Although similar trends are observed in detection performance, the analytical framework developed in this study is based on a more comprehensive and unified approach, thus, better suited for the purpose of optimizing biosensors from MC theoretical perspective.

The remainder of the paper is organized as follows. In Section II, we describe bioFETs and explain their operation principles. We develop the model of SiNW FET-based MC receivers in Section III. In Section IV, we derive the SEP for an MC system that employs the nanobioelectronic receiver and utilizes M-CSK scheme for modulation. The performance evaluation results are presented in Section V. Finally, the concluding remarks are given in Section VI.
\begin{figure}[!t]
\centering
\includegraphics[width=9cm]{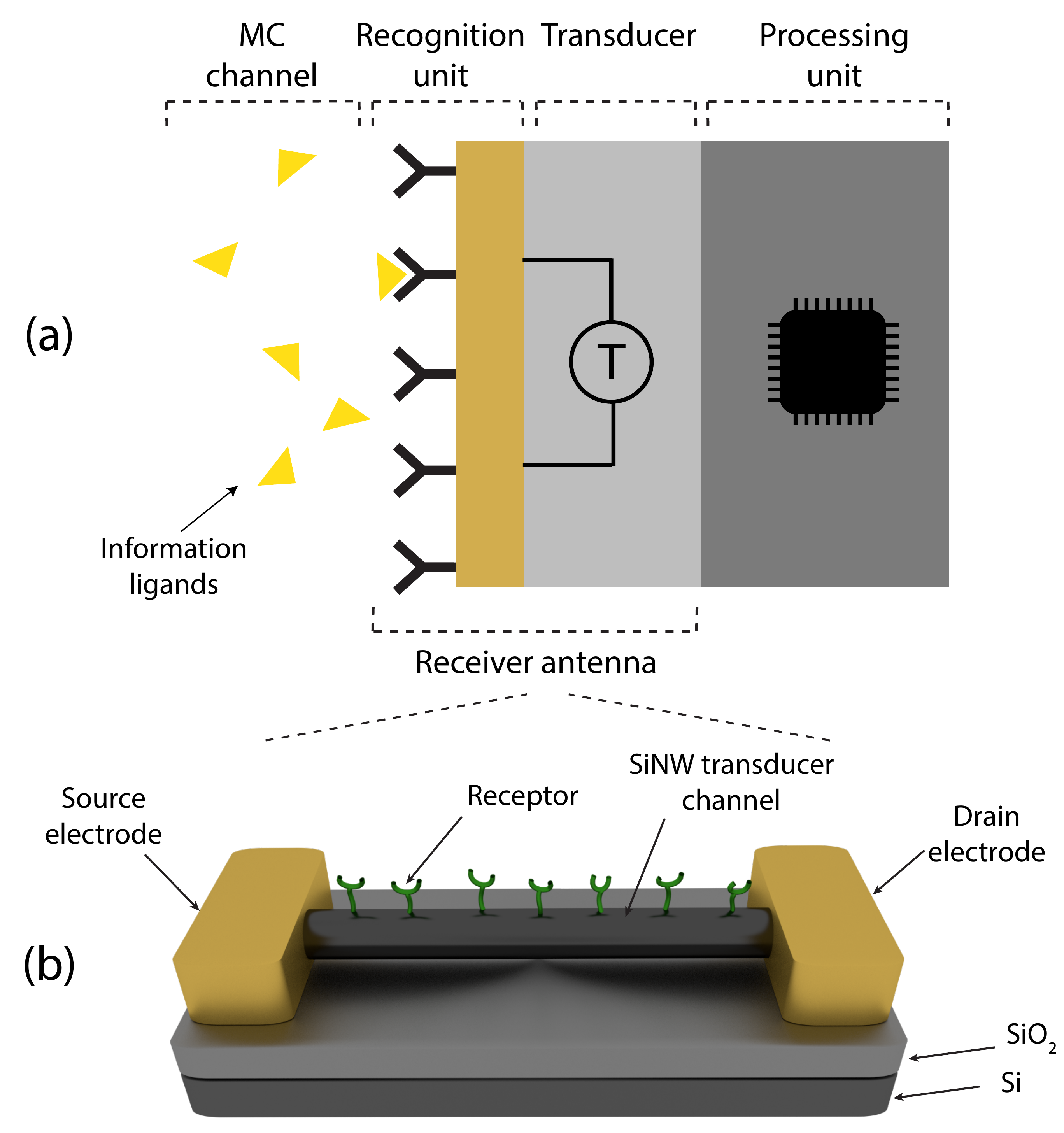}
\caption{(a) Functional units of an MC receiver, and (b) SiNW FET-based MC receiver antenna. Insulating SiO$_2$ layer entirely covering the SiNW is not shown in (b) for better visualization of the transducer.}
\label{fig:MCreceiver}
\end{figure}

\section{Principles of BioFETs}
Operation principles of bioFET, which is the basis of the proposed MC receiver architecture, are similar to the ones of the conventional FETs. In conventional FET type transistors, current flows from the source electrode to the drain electrode through a semiconductor channel, conductance of which is controlled by the electric field created by the potential applied on the gate electrode. Conductivity is proportional to the density of the carriers accumulated in the channel, and the variations of the electric field resulting from the additional surface potential is reflected to the changes in the voltage-current characteristics between the drain and the source electrodes.

BioFETs slightly differ from conventional FETs by including a biorecognition layer capable of selectively binding target molecules \cite{Poghossian2014}. This layer is composed of receptor molecules tethered on the surface of the FET channel, and replaces the gate electrode of conventional FETs, as shown in Fig. \ref{fig:MCreceiver}. Binding of ligands to the surface receptors results in accumulation or depletion of carriers in the semiconductor channel due to the field effect generated by the intrinsic charges of the bound ligands. Hence, the ligand binding modulates channel conductance and current.

Several ligand-receptor pairs, e.g., antibody-antigen, aptamer-natural ligand, natural receptor/ligand, have proven suitable for bioFETs \cite{Rogers2000}. Various types of semiconductors, such as SiNW, Carbon NanoTube (CNT) and graphene, can be used as the FET channel, i.e., transducer channel \cite{Poghossian2014}. The basics of biorecognition and transducing operations and the noise processes do not fundamentally differ based on the type of the ligand-receptor pair and the semiconductor channel. However, the literature is currently dominated by the studies focusing on SiNW bioFETs. One inherent drawback of SiNW bioFETs is the thin oxide layer built up around the SiNW surface, which deteriorates its detection performance \cite{Senveli2013}. Therefore, researchers are in search for other nanomaterials suitable for application in biosensing. For example, metal-oxide nanowires, such as SnO$_2$, ZnO, In$_2$O$_3$, are rapidly proving to be useful in nanoscale biosensing applications as the active channels of thin film biotransistors \cite{Huang2009}. However, their fabrication is currently constrained by bottom-up methods, which limits the controllability of their size and doping characteristics \cite{Senveli2013}. In contrast, SiNW provides a wider range of fabrication options including lithographic top-down methods. Together with a more established literature, easier and controllable fabrication of SiNWs lead us to focus on SiNW bioFETs in this study to develop a model for bioFET-based MC receiver.

\section{SiNW FET-Based Receiver Model}
\subsection{Model Description}
We consider a time-slotted molecular communication system between a single transmitter receiver pair, which are assumed to be perfectly synchronized with each other in terms of time. The system utilizes M-ary concentration shift keying (M-CSK) modulation such that information is encoded into the concentration, i.e., the number, of molecules. Given that the input alphabet is $\mathcal{M} = \{0, 1,...,M-1\}$, to send the symbol $m \in \mathcal{M}$ for the $k^{th}$ time slot, the transmitter releases $N_m$ molecules at the beginning of the signaling interval, i.e., at time $t_k=kT_s$, where $T_s$ is the slot duration, i.e., the symbol period.

\begin{figure*}[!t]
\centering
\includegraphics[width=16cm]{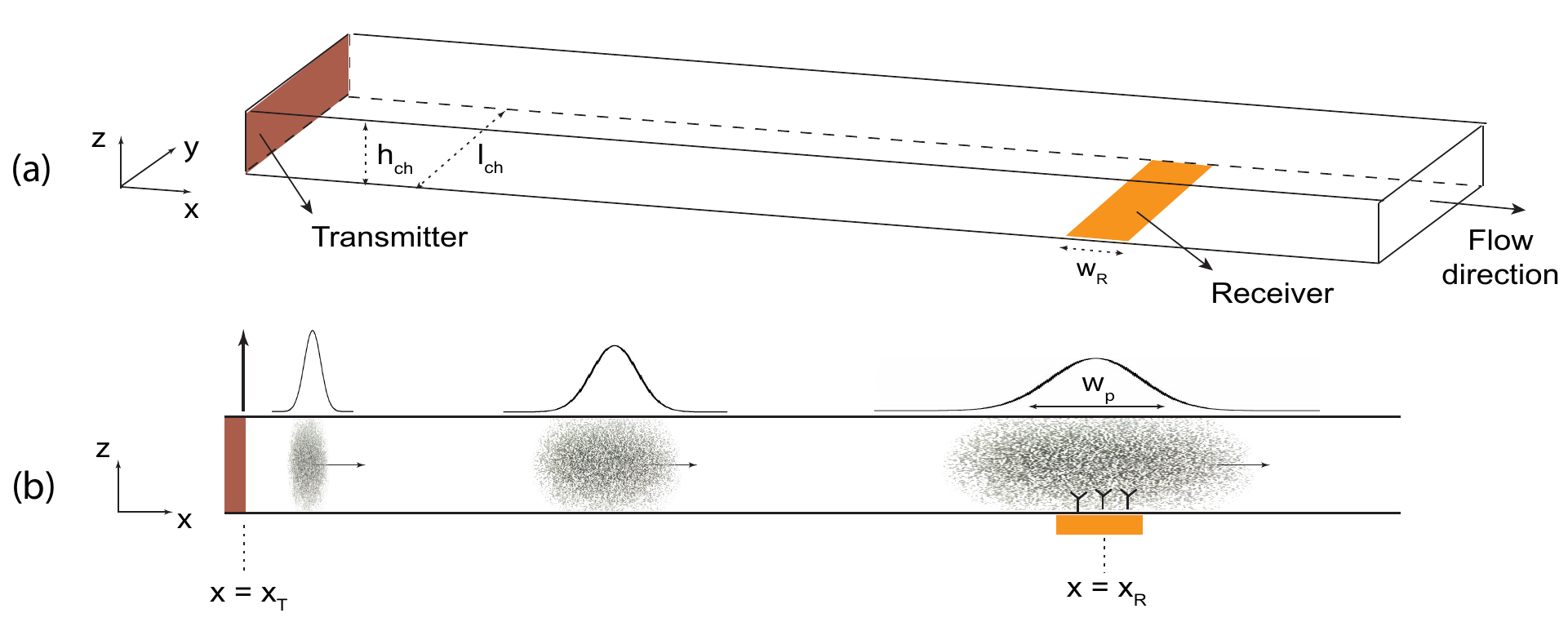}
\caption{(a) 3D and (b) 2D dimensional view of microfluidic propagation channel. The locations of the transmitter and receiver, and the plug of ligands, which is dispersed as it flows towards the receiver location, are shown. }
\label{fig:microfluidic}
\end{figure*}

For the propagation medium, we consider a simple straight microfluidic channel with a rectangular cross-section, as shown in Fig. \ref{fig:microfluidic}. The transmitter is assumed to be located at the entrance of the channel. A SiNW FET-based MC receiver is considered to be located at the bottom of the microfluidic channel at position $x = x_R$, with its SiNW transducing channel covered by the oxide layer and surface receptors, which are directly exposed to the information molecules, i.e., ligands, of varying concentration. The receiver samples the concentration of ligands flowing over its surface based on ligand-receptor binding kinetics \cite{Pierobon2011}. A fluid flows unidirectionally from the transmitter to the receiver location in the microfluidic channel such that the transmitted ligands are propagated to the receiver through advection and diffusion. Trajectories of each molecule along the propagation channel are assumed to be independent of each other.

In the considered scenario, ligands are not absorbed by the receiver, instead they temporarily bind to the surface receptors and unbind after a random amount of time. This characteristic of the ligands together with their long propagation times and spread over the x-axis through diffusion make the propagation channel have a memory, that may result in intersymbol interference (ISI) \cite{Bicen2015}. ISI can be overcome by selecting the symbol period $T_s$ sufficiently long, or employing auxiliary enzymes in the channel that degrade the information molecules in the environment after the detection, as proposed in \cite{Noel2014b}. To simplify the derivation of the receiver model, without loss of generality, we assume that the propagation channel is memoryless, thus, we neglect ISI.

\begin{figure*}[!t]
\centering
\includegraphics[width=18cm]{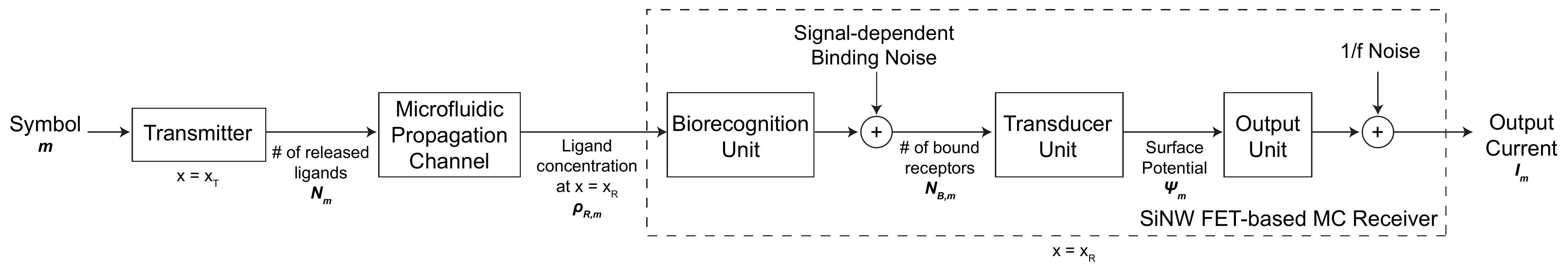}
\caption{Block diagram of microfluidic MC system with SiNW FET-based MC receiver.}
\label{fig:block}
\end{figure*}

The block diagram of the communication system including the SiNW FET-based MC receiver is shown in Fig. \ref{fig:block}. The receiver's operation can be described by the operations of three consecutive units. The Biorecognition Unit (BU) constitutes the interface of the receiver with the communication channel and is responsible for selectively sensing the concentration of ligands. In the Transducer Unit (TU), the ligands, which stochastically bind the surface receptors, modulate the gate potential of the FET through the field effect resultant from their intrinsic charges. In the Output Unit (OU), the modulated gate potential is immediately reflected into the current flowing through the SiNW channel between the drain and the source electrodes of the FET.

\subsection{Molecular Transport in Microfluidic Channel}
Transport dynamics of ligands inside the microfluidic channel can be described by the advection-diffusion equation. The fluid flow, which may be created by a pressure difference between the two ends of the channel, is taken as laminar, steady, and unidirectional along the channel's longitudinal axis, i.e., x-axis \cite{Bicen2013}. Assuming that a relatively low number of ligands reversibly react with the surface receptors and do not substantially change the concentration in the channel, the propagation of the ligands throughout the channel is handled as a one-dimensional advection-diffusion problem such that the ligand concentration $\rho$ and the fluid velocity $u$ are represented by their average over the channel's cross section \cite{Karsenty2014}. Therefore, the concentration and fluid flow are invariant along the y- and z-axis. One dimensional advection-diffusion equation along the direction of the fluid flow, $\vec{x}$, can be written as
\begin{equation}
\frac{\partial \rho(x, t)}{\partial t} = D \frac{\partial^2 \rho(x, t)}{\partial x^2} - u \frac{\partial \rho(x, t)}{\partial x}, \label{convection-diffusion}
\end{equation}
where $\rho(x, t)$ is the ligand concentration at position $x$ and time $t$, and $u$ is the x-axis fluid flow velocity averaged over the channel's cross section. $D$ is the effective diffusion coefficient that accounts for the effect of Taylor-Aris type dispersion \cite{Bicen2013}. For a channel with rectangular cross-section, it is given by
\begin{equation}
D = \left(1 + \frac{8.5 u^2 h_{ch}^2 l_{ch}^2 }{210 D_0^2 (h_{ch}^2 + 2.4 h_{ch} l_{ch}+ l_{ch}^2)} \right) D_0, \label{ficks}
\end{equation}
where the intrinsic diffusion coefficient is denoted by $D_0$ \cite{Bicen2013}. $h_{ch}$ and $l_{ch}$ are the cross-sectional height and length of the channel, respectively. Transmitter is assumed to be a planar ligand source located at $x = x_T$ and release a preset number of ligands $N_m$ at time $t_k=k T_s$ into the channel for representing symbol $m$. Since we neglect ISI, to model the propagation, it is sufficient to consider only one signaling interval, say $k = 0$ and $t_k = 0$. Taking the transmitter location as $x_T = 0$, and assuming that ligands are uniformly distributed over the channel's cross-section at the release time, the initial condition can be given as an impulse scaled by surface concentration:
\begin{equation}
\rho_m(x, t = 0) = \frac{N_m}{A_{ch}} \delta(x), \label{initial}
\end{equation}
where $A_{ch} = h_{ch} \times l_{ch}$ is the cross-sectional area of the channel, and $\delta$ is the Dirac delta function. Accordingly, the solution of \eqref{convection-diffusion} for $t > 0$ is given by \cite{Bicen2013} as follows
\begin{equation}
\rho_m(x, t) = \frac{N_m / A_{ch}}{\sqrt{4 \pi D t}} \exp\left(-\frac{(x- u t)^2}{4 D t} \right). \label{cd-solution}
\end{equation}

\subsection{Received Signal}
The SiNW FET-based receiver is considered to be placed at the bottom of the microfluidic channel and located along the transverse axis perpendicular to fluid flow. We assume that the length of the SiNW, $l_R$, is equal to the cross-sectional length of the channel, i.e., $l_R = l_{ch}$, and the drain, source and gate electrodes of the receiver are buried inside the channel walls and do not affect the fluid flow and ligand propagation. The radius of the SiNW is denoted by $r_R$, and the position of the radial axis of the SiNW is taken as the receiver's center position, i.e., $x_R$.

As can be inferred from \eqref{cd-solution}, the ligand concentration profile is Gaussian at any observation time. As the ligands are transported by the fluid flow along the channel, they are dispersed and the peak ligand concentration gets attenuated. We define the propagation delay as the expected time it takes for the peak concentration to arrive at the center position of the receiver $x=x_R$:
\begin{equation}
t_D = \frac{x_R}{u}. \label{delay}
\end{equation}
Assuming that the transmitter and receiver are perfectly synchronized in time, the receiver is supposed to sample the receptor states at time $t = t_D$. We assume that all of the points on the receiver surface are exposed to the same concentration.

The ligands flowing over the receiver surface interact with the surface receptors through ligand-receptor binding mechanism. To make the derivation of binding probability and the statistics of the binding fluctuations in Section \ref{biorecognition} analytically tractable, we also assume that the surface receptors are at equilibrium with a stationary ligand concentration at the time of sampling. The reasoning is that the advection-diffusion channel behaves like a low-pass filter \cite{Bicen2013}, which makes the concentration signal at the receiver location slowly varying, allowing the receptors with relatively fast kinetics reach the equilibrium state. Specifically, we consider that $\rho_m(x_R,t)$ is fixed to its peak value $\rho_m(x_R, t_D)$ when it takes values from the interval $[0.99 \rho_m(x_R, t_D), \rho_m(x_R, t_D)]$. So we neglect the small variations limited to 1$\%$ of the peak concentration at the received signal, assuming that they do not disturb the equilibrium. This portion of the ligand concentration signal completes its passage over the receiver surface in an approximate duration of
\begin{equation}
\tau_p = \frac{4}{u} \sqrt{-\ln(0.99) D t_D}. \label{duration}
\end{equation}
In Section \ref{biorecognition}, we will use $\tau_p$ to set a constraint for the kinetic rates of ligand-receptor binding reaction to satisfy the equilibrium assumption. In line with the above assumptions, the input signal of interest for the receiver can be written as
\begin{equation}
\begin{split}
\rho_m(x_R,t) &\approx \rho_m(x_R,t_D) = \frac{N_m}{A_{ch} \sqrt{4 \pi D t_D}}, \\
& ~~\text{for}\ t \in [t_D-\tau_p/2, t_D+\tau_p/2] . \label{received}
\end{split}
\end{equation}
For ease of notation, we use $\rho_{R,m} = \rho_m(x_R,t_D)$ to denote the received molecular signal.

\subsection{Biorecognition Block and Binding Noise}
\label{biorecognition}
\begin{figure}[!t]
\centering
\includegraphics[width=9cm]{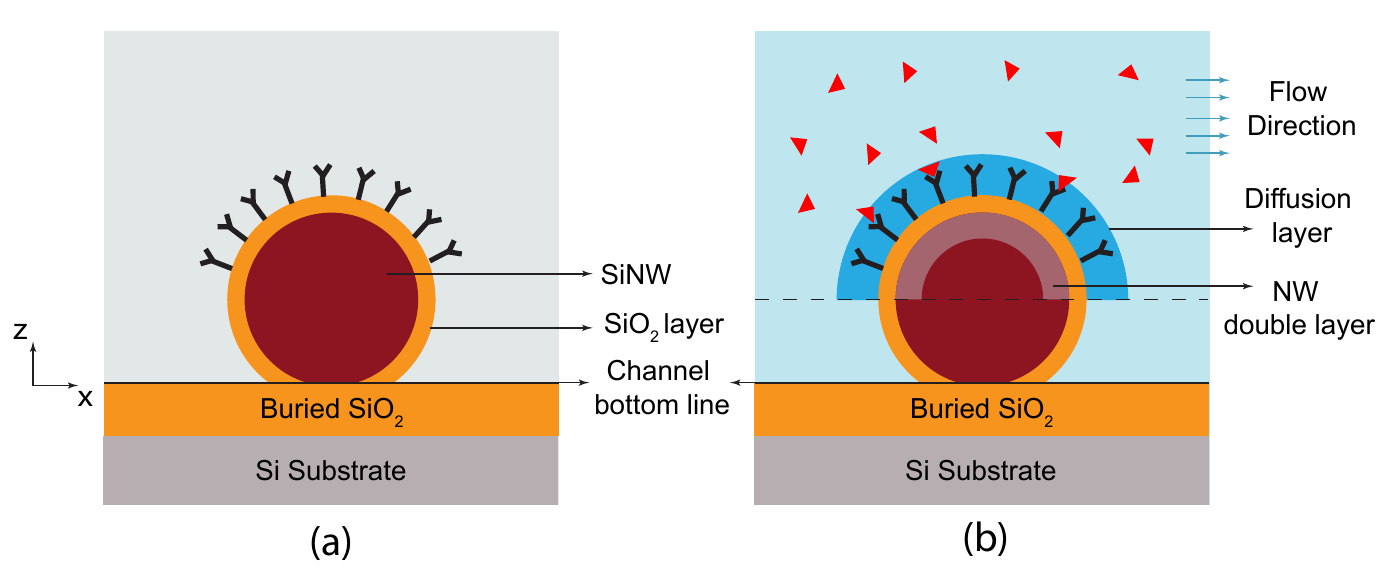}
\caption{Cross-sectional view of SiNW FET-based MC receiver buried at the bottom of the microfluidic channel, (a) when the channel is empty; (b) when the receiver is exposed to the electrolyte solution with ligands flowing along the x-axis. We show only the receptors at the top surface of the SiNW, since we assume that the ligands can bind only to these receptors. The top part of the diffusion layer resulting from the counterions attracted to the insulating SiO$_2$ layer, and the corresponding SiNW double layer emerged inside the SiNW are also shown in (b). The region of interest is the zone lying above the dashed line, which is the effective part of the receiver for ligand detection.}
\label{fig:SiNW}
\end{figure}

We start modeling the biorecognition block by first investigating the ligand flux to the receiver surface. Since the SiNW resides on the surface of bulk SiO$_2$, which occludes the ligand flux from the bottom, ligands can be assumed to interact only with the receptors tethered to the top surface, as shown in Fig. \ref{fig:SiNW}. Thus, we can consider that the receiver has a hemicylindrical interface to the fluid. The transport rate, i.e., flux, of ligands to a hemicylindrical surface, placed at the bottom of a microfluidic channel with rectangular cross-section is investigated in \cite{Sheehan2005}. The authors show that it can be approximated (with an error of $\sim 5\%$) as \cite{Sheehan2005}:

\small
\begin{multline}
k_T = D l_r \times \\
\begin{cases}
     \left(0.8075 P_s^{1/3} + 0.7058 P_s^{-1/6} - 0.1984 P_s^{-1/3}\right),& \text{if } P_s > 1\\
     \frac{2 \pi}{4.885 - \ln{(P_s)}} \left( 1 - \frac{0.09266 P_s}{4.885 - \ln{(P_s)}} \right), & \text{if } P_s < 1
\end{cases} \label{transportrate}
\end{multline}
\normalsize
where $P_s = (6 Q w_R^2)/(D l_{ch} h_{ch}^2)$, with $Q = u \times A_{ch}$ being the volumetric flow rate and $w_R =  \pi r_R$ being the effective width of the SiNW.

In line with the assumptions made for the ligand transport, the mean number of bound receptors on the receiver surface $\mu_{N_{B,m}}(t)$ is governed by the following differential equation
\begin{equation}
\frac{d\mu_{N_{B,m}}(t)}{dt} = k_1^* \rho_m(x_R, t) \left[N_R - \mu_{N_{B,m}}(t) \right] - k_{-1}^* \mu_{N_{B,m}}(t), \label{diff1}
\end{equation}
where $N_R$ is the number of surface receptors (assumed to be non-interacting with each other), and $\rho_m(x_R, t)$ is the solution \eqref{cd-solution} at $x = x_R$. Here $k_1^* = k_T k_1/(k_T + k_1)$ and $k_{-1}^* = k_T k_{-1}/(k_T + k_{1})$ are the effective binding and unbinding rates, which incorporate the effect of transport rate $k_T$ to the intrinsic binding rate $k_1$ and unbinding rate $k_{-1}$ of the ligand-receptor pair \cite{Berezhkovskii2013}. Note that $k_T$ has no effect on the equilibrium dissociation constant $K_D = k_{-1}/k_{1} = k_{-1}^*/k_{1}^*$. Equation \eqref{diff1} does not yield an analytical solution for $\mu_{N_{B,m}}$; however, with the assumptions of stationary ligand concentration fixed to $\rho_{R,m}$ and reaction equilibrium, the solution at the sampling time $t_D$ can be given by
\begin{equation}
\mu_{N_{B,m}} = P_{on|m} N_R = \frac{\rho_{R,m}}{\rho_{R,m}+K_D} N_R, \label{Nbmean}
\end{equation}
where $P_{on|m}$ is the probability of finding a single receptor in the ON (bound) state at equilibrium \cite{Berezhkovskii2013}. The equilibrium probability distribution $P(N_B = n)$ of having $n$ number of bound receptors is a Binomial distribution \cite{Berezhkovskii2013}, with the variance of
\begin{equation}
\sigma_{N_{B,m}}^2 = P_{on|m} (1-P_{on|m}) N_R. \label{Nbvar}
\end{equation}

The correlation time of the fluctuations in the number of bound receptors at equilibrium, which we call \emph{the binding noise}, is determined by the kinetic rate constants of the surface reaction, as well as influenced by the ligand transport rate $k_T$. It is equal to the relaxation time, i.e., mixing time, of transport-influenced ligand-receptor binding reaction, and given by \cite{Berezhkovskii2013} as follows
\begin{equation}
\tau_{B,m} = \frac{1}{k_1 \rho_{R,m} + k_{-1}} + \frac{k_1 (k_1 \rho_{R,m} + N_R k_{-1})}{ k_T (k_1 \rho_{R,m} + k_{-1})^2}. \label{correlation}
\end{equation}
For high values of $k_T$, i.e., in the reaction-limited case, the second term in the RHS of \eqref{correlation}, which incorporates the effect of ligand transport rate, would vanish; and in this case, the correlation would be governed only by the reaction terms.
Following \cite{Berezhkovskii2013}, the autocorrelation function (ACF) of binding noise is approximated with a single exponential as follows
\begin{multline}
R_{N_{B,m}}(t, t+\tau) = R_{N_{B,m}}(\tau) \simeq \sigma_{N_{B,m}}^2 e^{-\frac{\tau}{\tau_{B,m}}}, \\ ~~\text{for}\ t, t+\tau \in [t_{D}-\tau_p/2, t_{D}+\tau_p/2], \label{acfbinding}
\end{multline}
The power spectral density (PSD) of the binding noise can then be found by taking the Fourier transform (FT) of the ACF \eqref{acfbinding}
\begin{equation}
S_{N_{B,m}}(f) = \mathcal{F}\left\{R_{N_{B,m}}(\tau)\right\} = \sigma_{N_{B,m}}^2 \frac{2 \tau_{B,m}}{1+(2 \pi f \tau_{B,m})^2}.
\end{equation}

For the surface receptors, the time to reach equilibrium is governed by the relaxation time $\tau_{B,m}$ given by \eqref{correlation}. It can be assumed that the receptors reach equilibrium when they expose to a stationary ligand concentration for a duration of time greater than $5 \tau_{B,m}$. Therefore, for the equilibrium assumption, we set a constraint on $\tau_p$, the time duration for which we assume the ligand concentration is constant (see \eqref{duration}), as follows
\begin{equation}
\tau_p \ge 5 \tau_{B,m}. \label{constraint}
\end{equation}
Given a ligand-receptor pair with particular binding/unbinding rates, this constraint sets a limit especially on the flow velocity $u$ and TX-RX distance $d$, which most affect the dispersion of ligands. Therefore, with this constraint, we avoid situations in which the ligand concentration is rapidly varying at the receiver location and the receptors do not have the sufficient time to reach the equilibrium. Note that for parameter values satisfying the constraint \eqref{constraint}, the error in the mean number of bound receptors resulting from the equilibrium assumption is calculated as less than $0.1 \%$ by comparing \eqref{Nbmean} with the numerical solution of \eqref{diff1} at the sampling time $t_D$.

\subsection{Transducer Block}
As the charged ligands bind to the surface receptors on the insulating SiO$_2$ layer, which covers the SiNW, the carrier distribution of SiNW near the SiO$_2$/SiNW interface is disturbed. The amount of ligand charge effective on the SiNW is given by $Q_m = N_{B,m} \, q_{eff} \, N_{e^-}$, where $N_{e^-}$ is the number of free electrons per ligand molecule. $q_{eff}$ is the mean effective charge that can be reflected to the NW surface by a single electron of a ligand molecule in the presence of ionic screening in the medium, which is also known as the Debye screening \cite{Stern2007}. The mean effective charge of a free ligand electron is degraded as the distance between the ligand electron and the transducer increases, and the relation is given by $q_{eff} = q \times exp(-r/\lambda_D)$, where $q$ is the elementary charge, and $r$ is the average distance of ligand electrons in the bound state to the transducer's surface \cite{Rajan2013}, which is assumed to be equal to the average surface receptor length, i.e., $r = l_{SR}$. The ionic strength of the medium is characterized by Debye length, $\lambda_D = \sqrt{(\epsilon_M k_B T)/(2 N_A q^2 c_{ion})}$, where $\epsilon_M$ is the dielectric permittivity of the medium, $k_B$ is the Boltzmann's constant, $T$ is the temperature, $N_A$ is Avogadro's number, and $c_{ion}$ is the ionic concentration of the medium \cite{Rajan2013}.

\begin{figure}[!t]
\centering
\includegraphics[width=7cm]{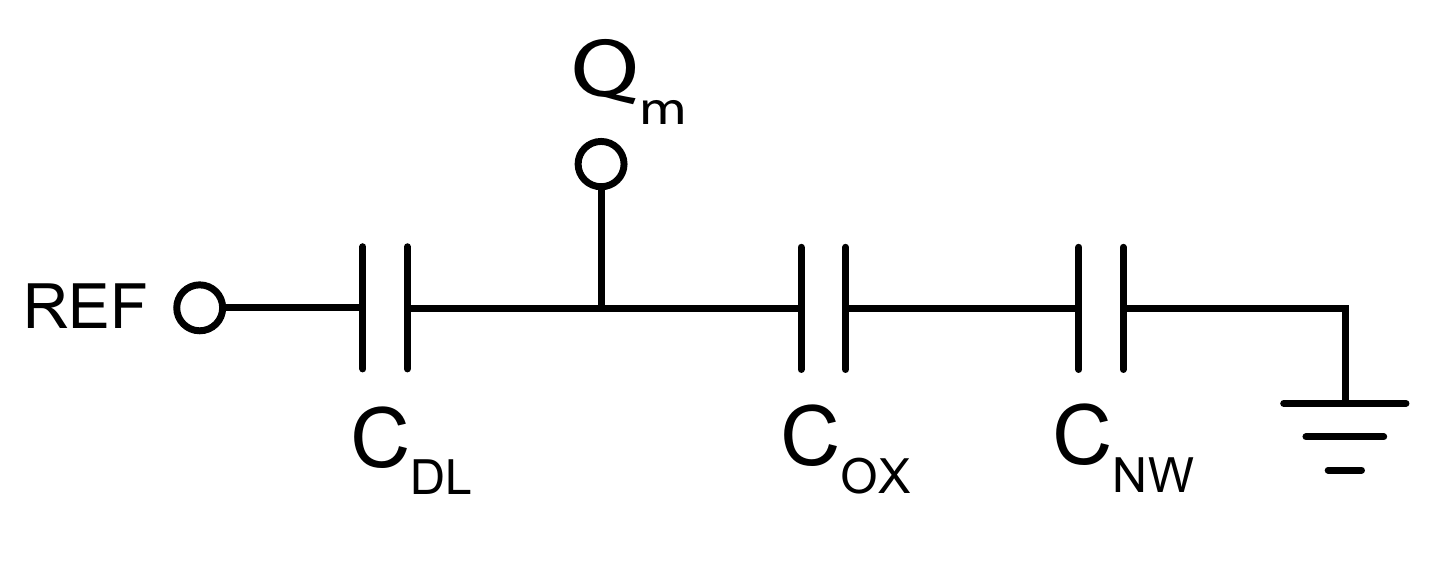}
\caption{Transducer equivalent circuit of the SiNW FET-based MC receiver \cite{Gao2010}, \cite{Shoorideh2014}. \emph{REF} denotes the reference electrode.}
\label{fig:Circuit}
\end{figure}

The ligand charges on the surface are transduced into a surface potential as follows \cite{Deen2006},
\begin{equation}
\Psi_m = \frac{Q_m}{C_{eq}}, \label{transduce}
\end{equation}
where $C_{eq}$ is the equivalent capacitance of the transducer. As demonstrated in Fig. \ref{fig:Circuit}, mainly three capacitances are effective on the transduction of the surface charges: (i) Diffusion layer capacitance, $C_{DL}$, resulting from the double layer created by the medium counterions accumulated at the interface between oxide layer, i.e., SiO$_2$ layer, and the electrolyte medium; (ii) the capacitance of the oxide layer; (iii) the SiNW capacitance, $C_{NW}$, which is again a double layer capacitance caused by the accumulation of carriers to the SiO$_2$/SiNW interface \cite{Gao2010}, \cite{Shoorideh2014}. Therefore, the equivalent capacitance can be given by
\begin{equation}
C_{eq} = \left(\frac{1}{C_{OX}} + \frac{1}{C_{NW}} \right)^{-1} + C_{DL}. \label{eq:debye}
\end{equation}
As we assume a nanowire-on-insulator (NWoI) configuration for the receiver design, individual capacitances can be obtained by considering the SiNW as a hemicylinder with an oxide layer of thickness $t_{OX}$ covering the SiNW, and a diffusion layer of thickness $\lambda_D$ covering the oxide layer \cite{Shoorideh2014}. Accordingly, the diffusion layer capacitance can be written as $C_{DL} = (\epsilon_{M}/\lambda_D) w_R l_R$. Similarly, the oxide layer capacitance can be given by $C_{OX} = (\epsilon_{OX}/t_{OX}) w_R l_R$, where $\epsilon_{OX}$ and $t_{OX}$ are the permittivity of the oxide layer. For high values of hole density, e.g., $p \sim 10^{18}$ cm$^{-3}$, corresponding to the linear operation regime of the FET, the double layer capacitance emerged in the NW channel can be given by $C_{NW} = (\epsilon_{Si}/\lambda_{Si}) w_R l_R$, where $\epsilon_{Si}$ is the dielectric permittivity of SiNW \cite{Gao2010}. $\lambda_{NW}$ is the thickness of the double layer created in the inner surface of the NW, which is given by $\lambda_{NW} = \sqrt{(\epsilon_{Si} k_B T)/(p q^2)}$.

\subsection{Output Block and $1/f$ Noise}
In the output block, the potential induced at the SiNW/oxide layer interface is reflected into a variation in the current flowing through the SiNW transducer channel. We assume that the p-type FET is operated in the linear region. Thus, the electrode voltage values satisfy
\begin{equation}
V_{SG} > |V_T|; ~~ V_{SD} \le V_{SG} - |V_T|, \label{muIx}
\end{equation}
where $V_{SG} = -V_{GS}$ is source to gate voltage, $V_{SD} = -V_{DS}$ is the source to drain voltage, and $V_T$ is the threshold voltage of the SiNW FET \cite{Sze2007}. The voltage values are assumed to be held constant during the receiver operation. In the linear region, the channel current is given by
\begin{equation}
I_{SD} =  \mu_p C_{OX} \frac{w_R}{l_R} \left[ \left(V_{SG} - |V_T|\right) V_{SD} - \frac{V_{SD}^2}{2} \right], \label{eq:Is}
\end{equation}
where $\mu_p$ is the carrier (hole) mobility, which depends on the impurity density of the transducer channel \cite{Sze2007}, and $l_R = l_{ch}$ as stated before. The partial derivative of the source-drain current with respect to source-gate voltage gives the transconductance of the FET \cite{Neamen2012}:
\begin{equation}
g_{FET} = \frac{\partial I_{SD}}{\partial V_{SG}} = \mu_p C_{OX} \frac{w_R}{l_R} V_{SD}, \label{eq:conductance}
\end{equation}
A potential created at the electrolyte/oxide layer interface due to the bound ligands acts upon the FET channel in the same way as the gate voltage does in a conventional FET. Therefore, the part of the channel current, $I_m$, generated by the surface potential, $\Psi_m$, can be written as
\begin{equation}
I_m = g_{FET} \times \Psi_m, \label{outputcurrent}
\end{equation}
Combining \eqref{received}, \eqref{Nbmean}, \eqref{transduce}, \eqref{outputcurrent}, mean of the generated output current can be given by
\begin{equation}
\mu_{I_m} = g_{FET}\Psi_L N_R \left( 1 + \frac{K_D A_{ch}}{N_m} \sqrt{ \frac{4 \pi D x_R}{u} }\right)^{-1}, \label{Imean}
\end{equation}
where $\Psi_L = (q_{eff} \times N_e^-)/C_{eq}$ is defined as the surface potential created by a single ligand.

As in all transistor devices, low-frequency operation of bioFET-based MC receiver is suffered from $1/f$ noise. In this paper, we use \emph{the correlated carrier number and mobility fluctuation model}, which provides an accurate description of the $1/f$ noise for FET type devices, attributing the source of $1/f$ noise to the random generation and recombination of charge carriers due to the defects and traps in the SiNW channel resulting from imperfect fabrication \cite{Sze2007}. The model expresses the resulting output current-referred noise PSD as
\begin{equation}
S_{I_m^F}(f) = S_{V,FB}(f) g_{FET}^2 \left[1 + \alpha_s \mu_p C_{OX} (V_{SG} - |V_{TH}|)  \right]^2 , \label{eq:Is}
\end{equation}
where $\alpha_s$ is the Coulomb scattering coefficient which depends on the temperature, and $\mu_p$ is the mobility of the hole carriers, which depends on the impurity concentration of the SiNW channel \cite{Rajan2010}. $S_{V,FB}$ is the PSD of the flatband-voltage noise:
\begin{equation}
S_{V,FB}(f) = \frac{\lambda k_B T q^2 N_{ot} g_{FET}^2}{w_R l_R C_{OX}^2 |f|},
\end{equation}
where $\lambda$ is the characteristic tunneling distance, $N_{ot}$ is the oxide trap density, i.e., impurity concentration, of the SiNW channel \cite{Rajan2010}. $1/f$ noise is independent of the received signals, and shows an additive behavior on the overall output current fluctuations \cite{Spathis2015}. Theoretically, $1/f$ noise does not have a low frequency cutoff; however, in experimental studies with a finite measurement time, a finite variance for $1/f$ noise is observed. The reason is related to the low frequency cutoff set by the observation time $T_{obs}$ \cite{Niemann2013}. Considering that the received molecular signals are at the baseband, to be able to calculate the total noise power, we assume one-year operation time, i.e., $ \sim \pi \times 10^{7}$ s, for the receiver such that the low cutoff frequency is $f_L = 1/T_{obs} \approx  1/\pi \times 10^{-7}$ Hz. At frequencies lower than $f_L$, the noise is assumed to show the white noise behavior, i.e., $S_{I_m^F}(f) = S_{I_m^F}(f_L)$ for $\lvert f \rvert < f_L$.

\subsection{Overall Noise PSD and Output SNR}
The PSD for the output current fluctuations due to the additive binding noise $S_{I_{m}^B}$ can be written as $S_{I_{m}^B}(f) = S_{N_{B,m}}(f) \times \Psi_L^2 \times g_{FET}^2$. Including the additive $1/f$ noise, the overall PSD of the output current referred noise is given by
\begin{equation}
S_{I_{m}}(f) = S_{I_{m}^B}(f) + S_{I_m^F}(f).
\end{equation}
Given the noise PSD, the output SNR of the receiver can be computed by
\begin{equation}
SNR_{out,m} = \frac{\mu_{I_{m}}^2}{\sigma_{I_m}^2}, \label{eq:SNRI}
\end{equation}
where $\sigma_{I_m}^2$ is the output current variance, obtained as follows
\begin{equation}
\sigma_{I_m}^2 = \int_{-\infty}^{\infty} S_{I_{m}}(f) df. \label{Ivar}
\end{equation}

\section{Gaussian Approximation and Symbol Error Probability}
\subsection{Gaussian Approximation for Noise Processes}
To analytically derive the SEP, we make reasonable approximations for the noise statistics. We can expect that a high number of receptors, e.g. $>1000$, are tethered to the top surface of a SiNW channel. Hence, the binomial distribution of the number of bound receptors can be approximated as Gaussian, i.e., $N_{B,m} \sim \mathcal{N}\left(\mu_{N_{B,m}}, \sigma_{N_{B,m}}^2\right)$, and the zero-mean additive binding noise follows normal distribution $\mathcal{N}\left(0, \sigma_{N_{B,m}}^2\right)$. The binding noise keeps Gaussian characteristics when passed through the linear filter in the transducing block.

$1/f$ noise is resulting from the bias current flowing through SiNW channel, therefore, it is independent of the binding noise. Although, there has been a long-standing discussion about the $1/f$ noise statistics, in many well-accepted experimental studies in the literature, it has been reported that $1/f$ noise can be approximated to follow a Gaussian distribution \cite{Bell1955}, \cite{Hooge1969}. In this paper, we rely on these reports to provide an analytical expression for the SEP.

Therefore, the overall noise process effective on the output current is the sum of two additive stationary noise processes that independently follow Gaussian statistics; thus, it is a stationary Gaussian process with a colored PSD.

\subsection{Derivation of SEP for M-CSK Modulation}
Let $H_m$ be the hypothesis that the symbol $m \in \mathcal{M}$ is transmitted at the beginning of $k^{th}$ time slot, and $Z_k$ be the output current sampled by the receiver for the $k^{th}$ slot. Then, with the Gaussian approximation of the additive noises, the conditional probability of $Z_k$ given that the hypothesis $H_m$ is true can be written as
\begin{equation}
P(Z_k|H_m) = \frac{1}{\sqrt{2 \pi \sigma_{I_m}^2}} e^{-\frac{(Z_k-\mu_{I_m})^2}{2\sigma_{I_m}^2}},
\end{equation}
where $\mu_{I_m}$ is the mean of the output current given by \eqref{Imean}, and $\sigma_{I_m}^2$ is the output current variance given by \eqref{Ivar}. Assuming that maximum likelihood (ML) detection is applied by the receiver, the decision rule can be expressed by
\begin{equation}
\hat{m}_k = \argmax_m P(Z_k|H_m),
\end{equation}
where $\hat{m}_k$ is the symbol decided at the receiver for the $k^{th}$ transmission. ML decision rule divides the entire range of the output current into $M$ decision regions corresponding to the $M$ symbols in the source alphabet. Decision region $D_m$ for the transmitted symbol $m$ can be defined as
\begin{equation}
\begin{split}
D_m = &\left\{ Z_k: P(Z_k|H_m) > P(Z_k|H_j), \forall j \neq m  \right\}, \\
& \text{for} ~~ m = 0,...,M-1. \label{threshold2}
\end{split}
\end{equation}
Assuming $N_0 < N_1 < ... < N_{M-1}$, from \eqref{Nbmean} and \eqref{Imean}, we know that the symbols satisfy the condition $\mu_{I_0} < \mu_{I_1} < ... < \mu_{I_{M-1}}$. Hence, the decision thresholds $\lambda_x$ separating the decision regions $D_{m-1}$ and $D_m$ can be obtained by comparing the conditional probabilities of adjacent symbols \cite{Singhal2015}
\begin{equation}
\begin{split}
\frac{1}{\sqrt{2 \pi \sigma_{I_m}^2}} e^{-\frac{(\lambda_m - \mu_{I_m})^2}{2 \sigma_{I_m}^2}} =&  \frac{1}{\sqrt{2 \pi \sigma_{I_{m-1}}^2}} e^{-\frac{(\lambda_m - \mu_{I_{m-1}})^2}{2 \sigma_{I_{m-1}}^2}}, \\ & \text{for} ~~ m = 1,...,M-1. \label{threshold}
\end{split}
\end{equation}
Solving we obtain the decision thresholds as
\begin{equation}
\begin{split}
& \lambda_m  = \frac{1}{{\sigma_{I_m}^2 - \sigma_{I_{m-1}}^2}} \Biggl( \sigma_{I_m}^2 \mu_{I_{m-1}} - \sigma_{I_{m-1}}^2 \mu_{I_m} \\
& + \sigma_{I_m} \sigma_{I_{m-1}} \sqrt{(\mu_{I_m} - \mu_{I_{m-1}})^2 + 2(\sigma_{I_m}^2 - \sigma_{I_{m-1}}^2) \ln{\frac{\sigma_{I_m}}{\sigma_{I_{m-1}}}}} \Biggr), \\
& \text{for} ~~ m = 1,...,M-1. \label{threshold2}
\end{split}
\end{equation}

The probability of erroneous detection can be computed as
\begin{equation}
P(e|H_m) = \int\limits_{z \notin D_m} P(z|H_m) dz,
\end{equation}
and the SEP can be given as follows
\begin{equation}
\begin{split}
P_e =& \frac{1}{M} \sum_{m=0}^{M-1} P(e|H_m) \\
    =& \frac{1}{2M} \Biggl[\erfc\left(\frac{\lambda_1-\mu_{I_0}}{\sigma_{I_0} \sqrt{2}}\right) + \erfc\left(\frac{\mu_{I_{M-1}}-\lambda_{M-1}}{\sigma_{I_{M-1}} \sqrt{2}}\right) \\
     &  + \sum_{m=1}^{M-2} \Biggl(\erfc\left(\frac{\mu_{I_m}-\lambda_{m}}{\sigma_{I_m} \sqrt{2}}\right) + \erfc\left(\frac{\lambda_{m+1}-\mu_{I_m}}{\sigma_{I_m}\sqrt{2}}\right)    \Biggr) \Biggr], \\
     & \text{for} ~~ m = 0,...,M-1, \label{decision2}
\end{split}
\end{equation}
where $\erfc(z) = \frac{2}{\sqrt{\pi}} \int_z^{\infty} e^{-y^2} dy$ is the complementary error function.

\section{Performance Analysis}
In this section, we present the numerical results obtained based on the developed model under different settings to reveal the performance of the SiNW FET-based MC receiver. The default values for the controllable parameters used in the analyses are listed in Table \ref{table:parameters}.

We assume that the microfluidic channel is filled up with an electrolyte, which is moderate in its ionic concentration (with $c_{ion}=30$ mol/m$^3$) compared to the diluted solutions (with $c_{ion}<1$ mol/m$^3$) used in \emph{in vitro} biosensing experiments \cite{Duan2012} and physiological solutions (with $c_{ion}>70$ mol/m$^3$) \cite{Okada1990}. The receptors are considered to be aptamers. The default length of receptors is set to 2 nm, which corresponds to 6 base pair-aptamers \cite{Kuscu2015b}. Binding and unbinding rates, $k_+$ and $k_-$, are set, considering the accepted values in the MC literature \cite{Pierobon2011} and the range of rates that aptamers can provide \cite{Luzi2003}. The relative permittivity of SiO$_2$ layer is reported as $\epsilon_{ox}/\epsilon_0 = 3.9$ \cite{Deen2006}. The thickness of the SiO$_2$ layer, $t_{ox}$, is a design parameter, for which we select a default value of $2$ nm. Depending on the fabrication, the tunneling distance for SiO$_2$ is on the order of 0.01-0.1nm \cite{Rajan2013}. We set $\lambda = 0.05$ nm as reported in \cite{Deen2006}. We assume the use of a p-type SiNW, which is moderately clean with the impurity density $N_{ot} = 10^{16}$ eV$^{-1}$cm$^{-3}$. This corresponds to a low-field mobility of $\mu_p = 500$ cm$^2$/Vs for hole carriers in SiNW (see Fig. 15 in \cite{Sze2007}). The Coulomb scattering coefficient is taken as $\alpha_s = 1.9 \times 10^{14}$ Vs/C, which is the value at $T =$ 300 K \cite{Rajan2010}. We consider a shallow microfluidic channel with cross-sectional height $h_{ch} = 3$ $\mu$m and length $l_{ch} = 15$ $\mu$m, resulting in a laminar and steady flow \cite{Bicen2013}, with average flow velocity of $u = 10$ $\mu$m/s, a moderate value widely observed in microfluidic literature \cite{Bicen2013}, \cite{Karsenty2014}.

\begin{figure}[!t]
\centering
\includegraphics[width=8cm]{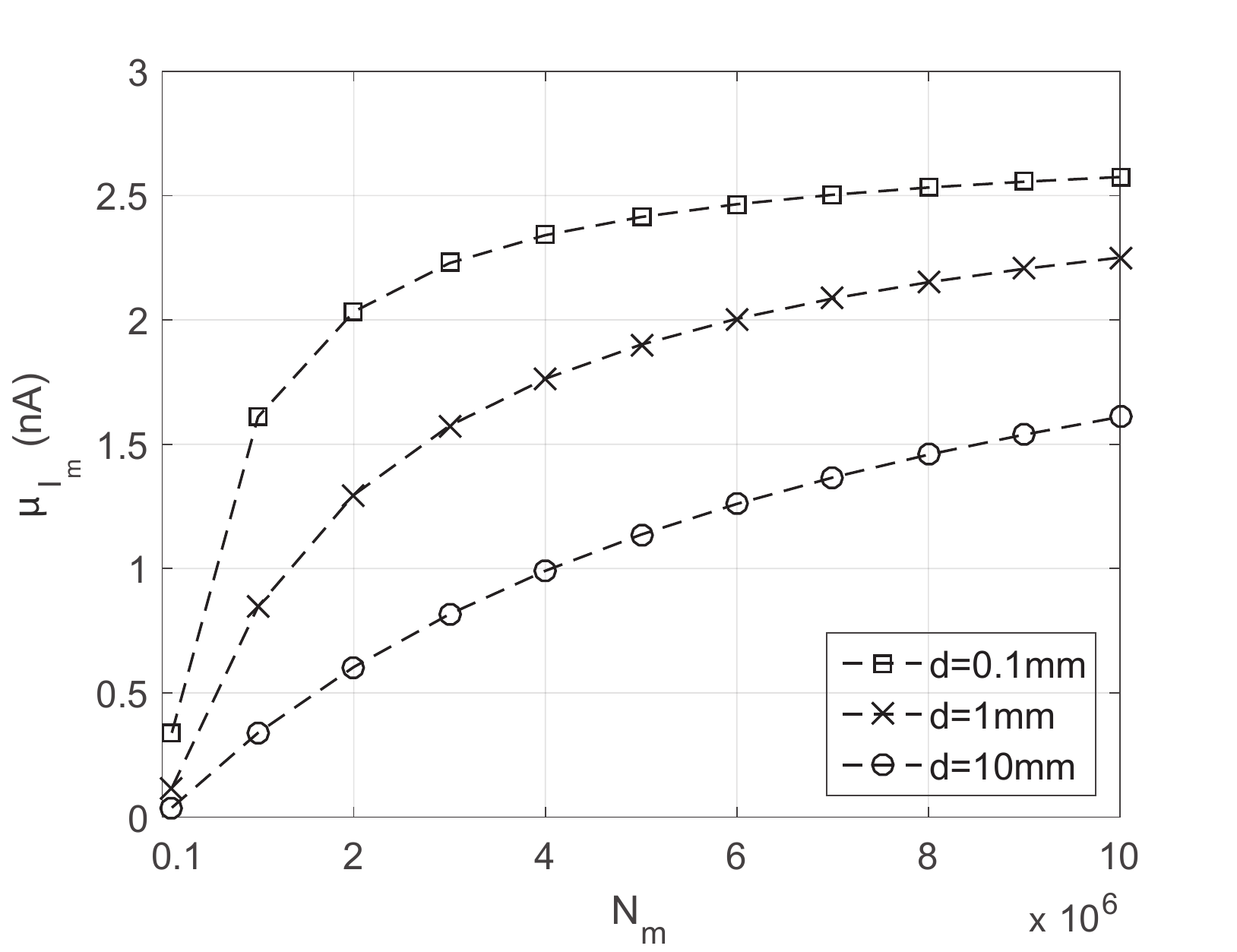}
\caption{Expected output current $\mu_{I_m}$ as a function of number of ligands $N_m$ released by transmitter and transmitter-receiver distance $d$.}
\label{fig:signal}
\end{figure}

\subsection{Receiver Response and Noise Power}
\subsubsection{Receiver Response}
We first investigate the expected response of the receiver to varying number $N_m$ of ligands released by the transmitter. The mean output current generated in the SiNW channel for several transmitter-receiver distances $d = x_R - x_T$ is plotted in Fig. \ref{fig:signal}. For each distance setting, we observe that the output current increases as the transmitter releases more ligands. However, at some value of $N_m$, the current begins to saturate. This is because as the receptors on the biorecognition layer are occupied by higher number of ligands; the receptors lose their sensitivity to the varying ligand concentration.
\begin{table}[!b]\scriptsize
\centering
\caption{Default Values of Simulation Parameters}
\begin{tabular}{ l | l }
   \hline \hline
   Microfluidic channel height ($h_{ch}$) & 3 $\mu$m  \\ \hline
   Microfluidic channel width ($l_{ch}$) & 15 $\mu$m  \\ \hline
   Number of transmitted ligands for symbol $m$ ($N_m$) & $5 \times 10^5$  \\ \hline
   Max number of ligands TN transmits ($K$) & $10^6$  \\ \hline
   Transmitter-receiver distance ($d$) & 1 mm  \\ \hline
   Average flow velocity ($u$) & 10 $\mu$m/s  \\ \hline
   Diffusion coefficient of ligands ($D_0$) & $2 \times 10^{-10}$ m$^2$/s  \\ \hline
   Binding rate ($k_1$) & $2 \times 10^{-19}$ m$^3$/s \\ \hline
   Unbinding rate ($k_{-1}$) & 20 s$^{-1}$  \\ \hline
   Average number of electrons in a ligand ($N_e^-$) & 3  \\ \hline  
   SiNW radius ($r_{R}$) & 10 nm  \\ \hline
   Concentration of receptors on the surface ($\rho_{SR}$) & $4 \times 10^{16}$ m$^{-2}$ \\ \hline
   Length of a surface receptor ($l_{SR}$) & 2 nm  \\ \hline   
   Temperature ($T$) & $300$K \\ \hline
   Relative permittivity of oxide layer ($\epsilon_{ox}/\epsilon_0$) & $3.9$  \\ \hline
   Relative permittivity of SiNW ($\epsilon_{NW}/\epsilon_0$) & $11.68$  \\ \hline
   Relative permittivity of medium ($\epsilon_R/\epsilon_0$) & $78$  \\ \hline
   Ionic strength of electrolyte medium ($c_{ion}$) & 30 mol/m$^3$  \\ \hline
   Source-drain voltage ($V_{SD}$) & $0.1$ V  \\ \hline
   Source-gate voltage ($V_{SG}$) & $0.4$ V  \\ \hline
   Threshold voltage ($V_{TH}$) & $0$ V  \\ \hline
   Hole density in SiNW ($p$) & $10^{18}$ $cm^{-3}$  \\ \hline
   Tunneling distance ($\lambda$) & $0.05$ nm  \\ \hline
   Thickness of oxide layer ($t_{ox}$) & $2$ nm  \\ \hline
   Oxide trap density ($N_{ot}$) & $10^{16}$ eV$^{-1}$cm$^{-3}$  \\ \hline
   Effective mobility of hole carriers ($\mu_{p}$) & $500$ cm$^2$/Vs  \\ \hline
   Coulomb scattering coefficient ($\alpha_s$) & $1.9 \times 10^{14}$ Vs/C  \\ \hline
\end{tabular}
\label{table:parameters}
\end{table}
\begin{figure}[!t]
\centering
\includegraphics[width=7.6cm]{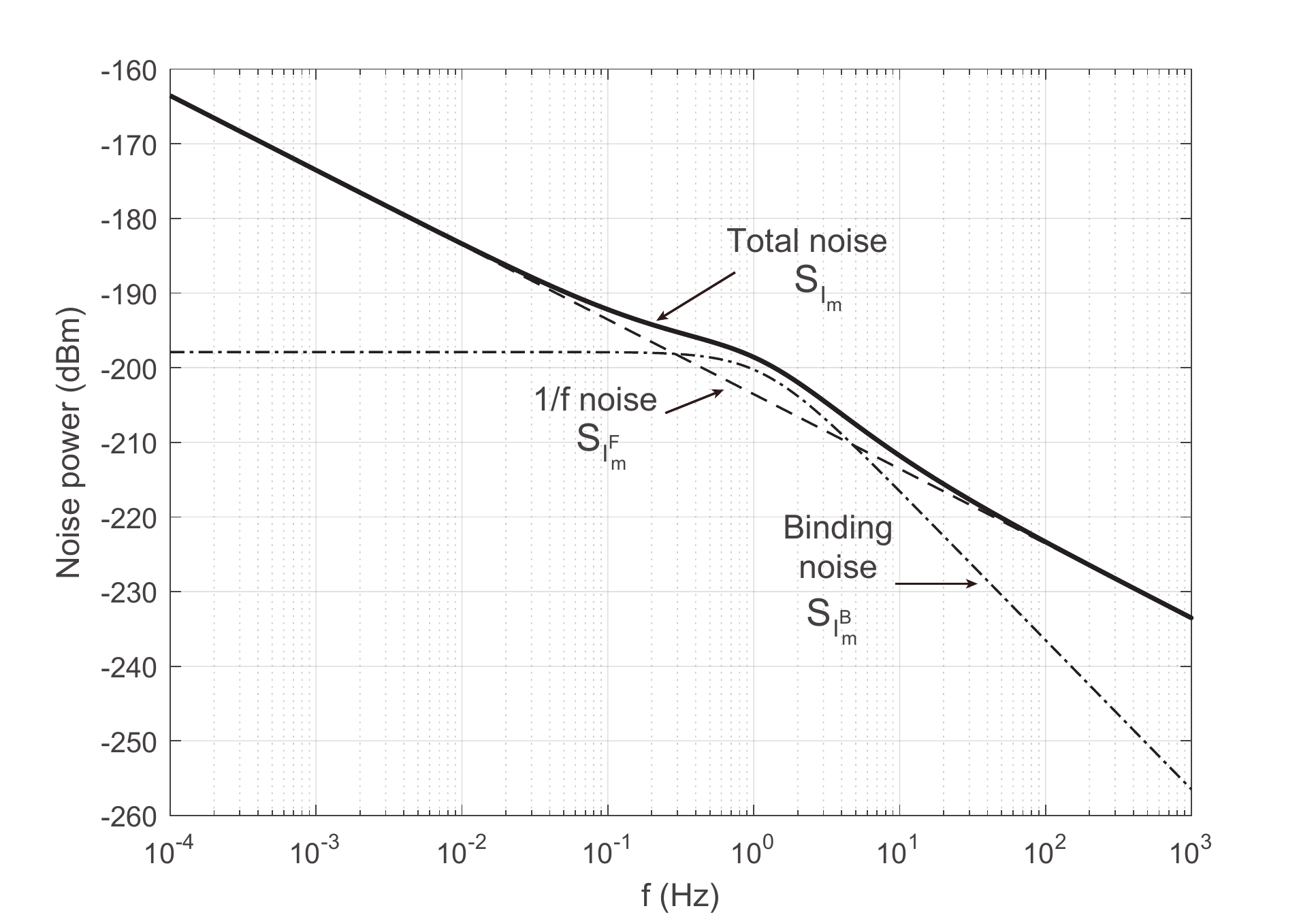}
\caption{PSD of noise effective on the output current of SiNW FET-based MC receiver. The plot reveals the individual contributions of binding and $1/f$ noise.}
\label{fig:PSD1}
\end{figure}
\begin{figure}[!t]
\centering
\includegraphics[width=8cm]{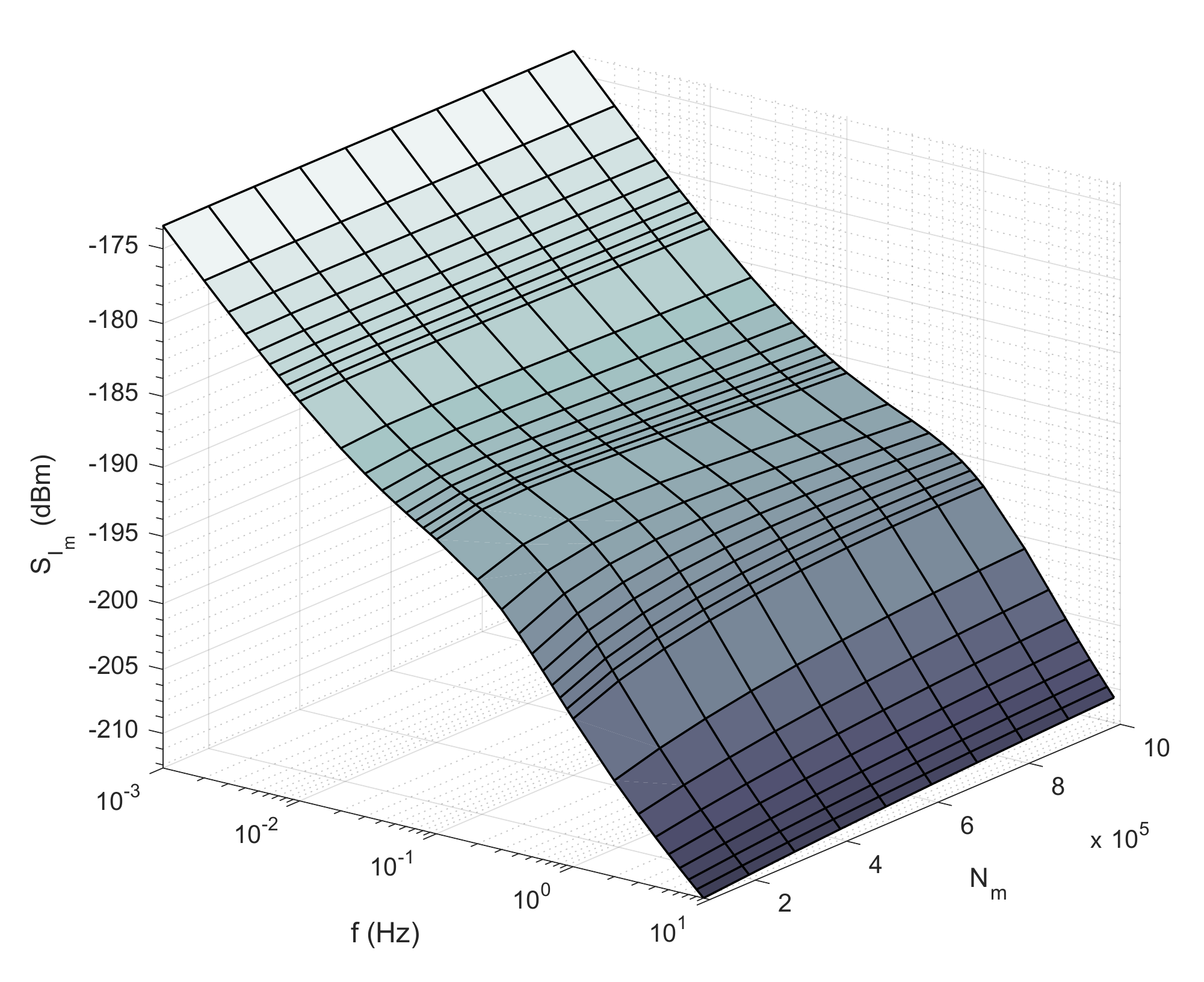}
\caption{Output current noise PSD at the frequency range of receiver's operation with varying $N_m$.}
\label{fig:PSD2}
\end{figure}
From the same figure, we can also infer that for the investigated range of number of released ligands, the receiver is most sensitive to the concentration variations when $d = 1$ mm. Since the attenuation of the concentration is proportional to $\sqrt{d}$ (see Equations \eqref{delay}, \eqref{received}); in the minimum distance case, i.e., when $d = 0.1$ mm, the ligand concentration observed at the receiver location is expected to be much higher. On the other hand, higher concentration of ligands in the receiver location leads to a more rapid saturation of the biorecognition unit, as evident from Fig. \ref{fig:signal}. When the distance is increased to $10$ mm, the ligand concentration over the receiver significantly decreases for the whole range of $N_m$, so that the variations of $N_m$ do not result in significant differences in the output current. Therefore, the dynamic range of the receiver and the attenuation of the molecular signals in the channel should be carefully considered while designing the overall MC system.

\subsubsection{Noise Power}
We plot the individual PSDs of binding and $1/f$ noises as well as the overall noise PSD effective on the output current. As seen in Fig. \ref{fig:PSD1}, the frequency domain is virtually divided into three regions, in each of which one of the two noise sources is prevailing. At very low frequencies, e.g., $f \ll 0.1$ Hz, $1/f$ noise is dominating over the binding noise, since the binding noise has a flat power density for frequencies below a critical value determined by the correlation time, i.e., $f_B = 1/\tau_B$, whereas the power of $1/f$ noise is increasing proportional to $1/f$. Around $f_B$, the binding noise may become dominant, depending on the total variance of the binding process. At frequencies higher than $f_B$, the binding noise is attenuated more rapidly than $1/f$ noise. Although $1/f$ noise is dominating again at high frequencies, its power decreases under $-220$ dBm; thus, the overall noise power is negligible in this frequency range.

The overall noise PSD is analyzed also for varying number of ligands released by the transmitter. As can be seen from Fig. \ref{fig:PSD2}, the contribution of $1/f$ noise dominating at very low frequencies does not vary as $N_m$ is changed; however, the contribution of the binding noise at low frequencies becomes more prevailing for lower ligand concentrations. This is because $\tau_B$ increases with decreasing ligand concentration, which decreases the critical frequency; see \eqref{correlation}.

\begin{figure*}[!t]
 \centering
   \subfigure[]{
 \includegraphics[width=4cm]{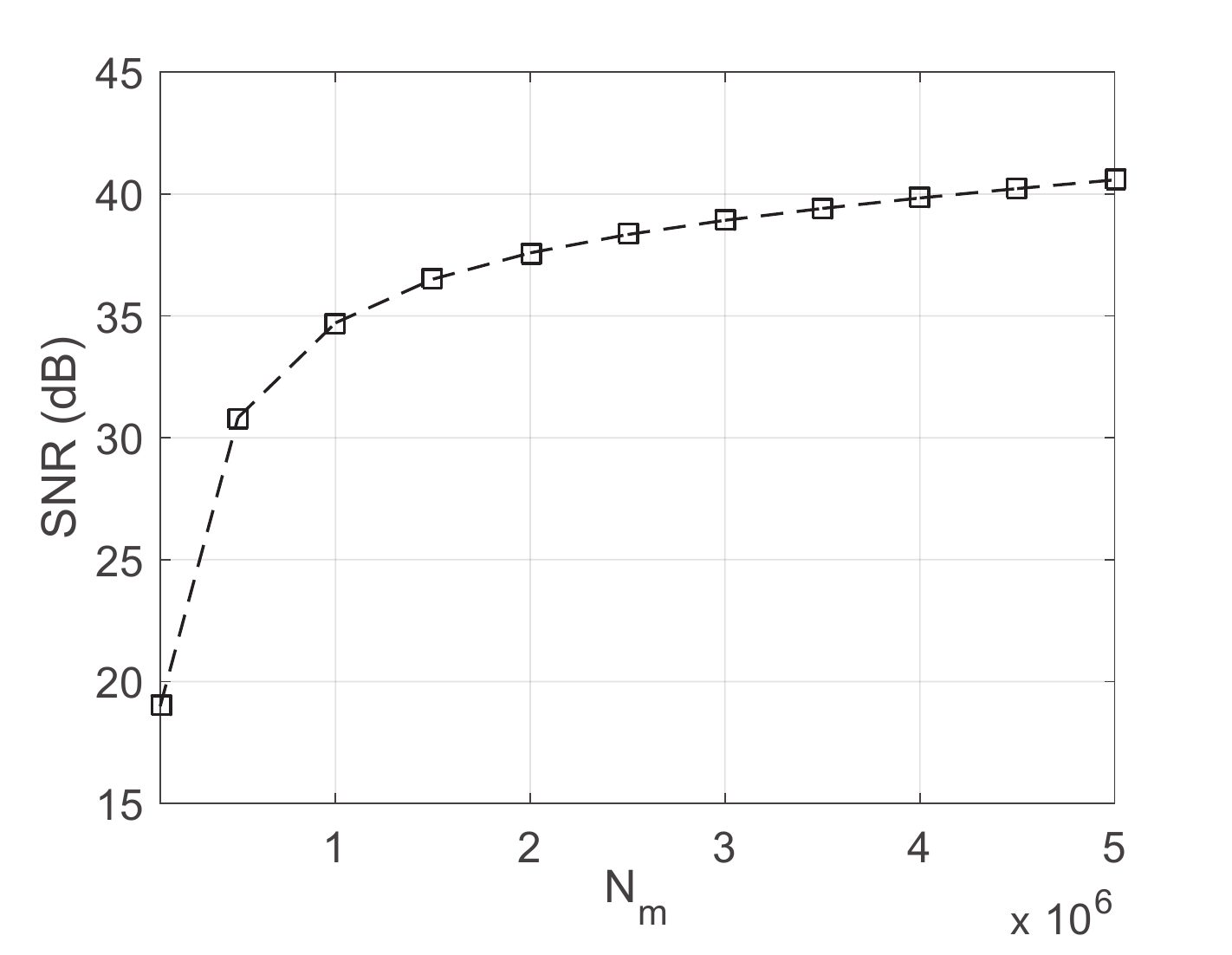}
 \label{fig:SNR_Nm}
 }
  \subfigure[]{
 \includegraphics[width=4cm]{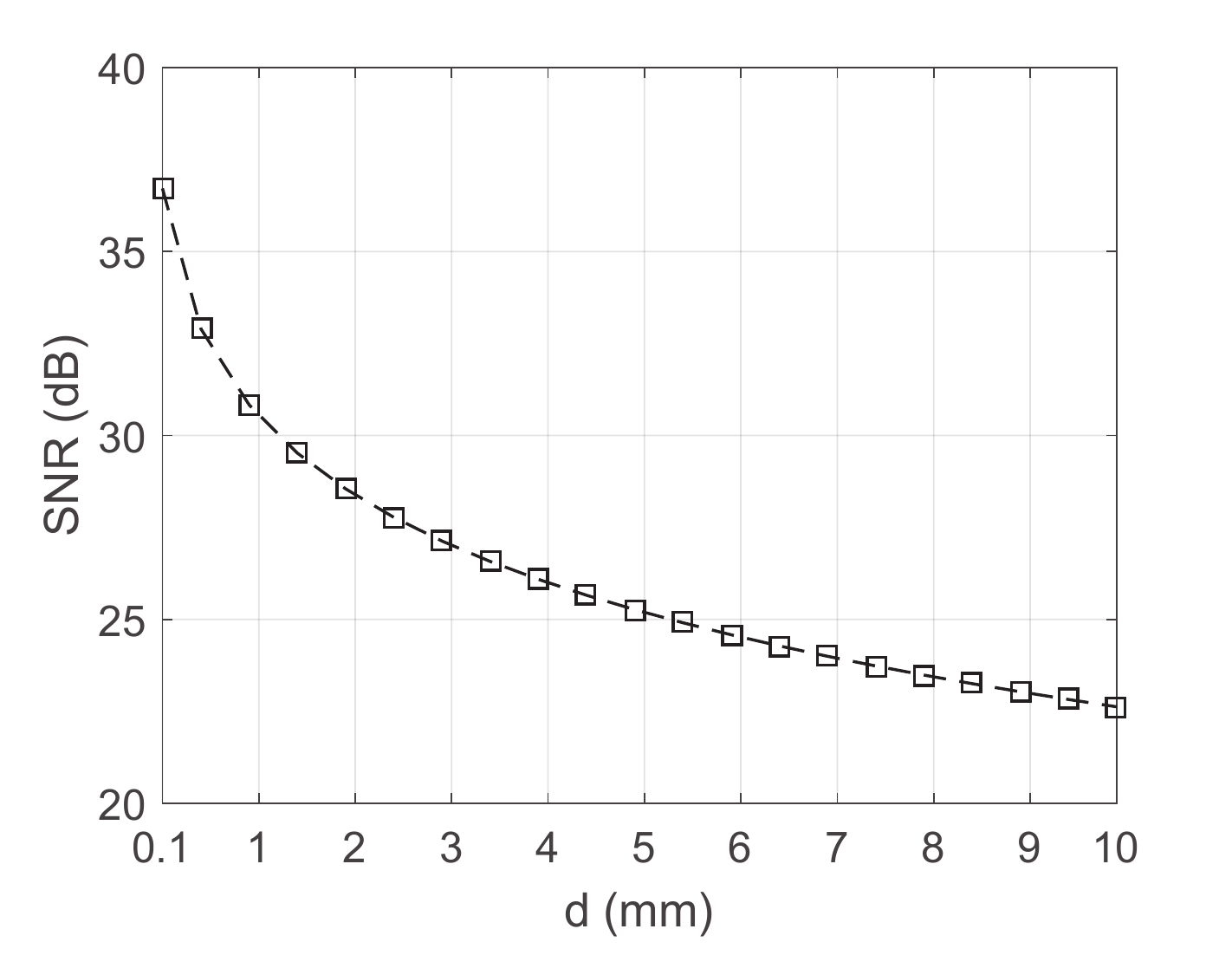}
 \label{fig:SNR_d}
 }
    \subfigure[]{
 \includegraphics[width=4cm]{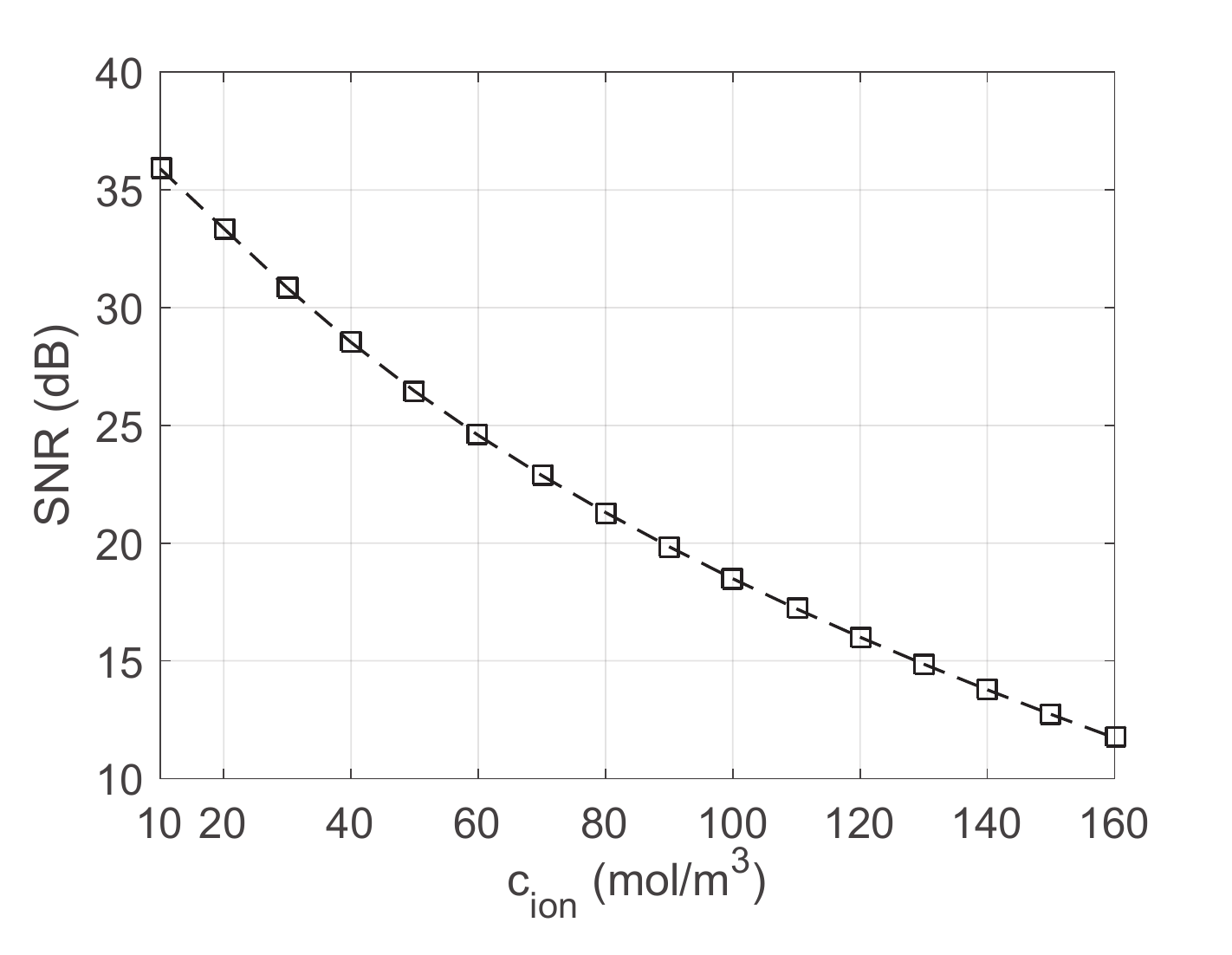}
 \label{fig:SNR_cion}
 }
  \subfigure[]{
 \includegraphics[width=4cm]{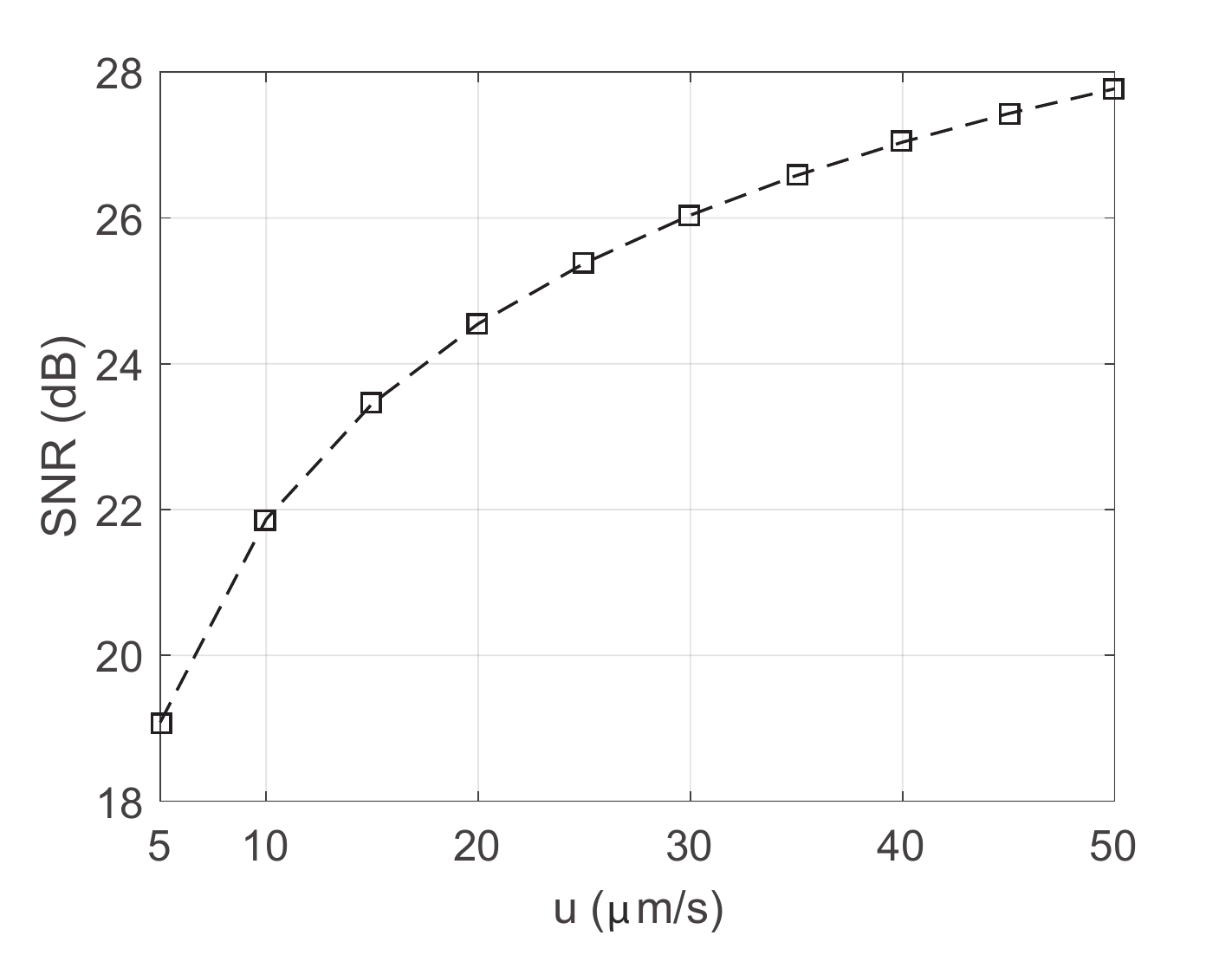}
 \label{fig:SNR_u}
 }
 \caption{Effect of the communication system parameters on the SNR at the electrical output of the receiver. SNR as a function of (a) number of transmitted ligands $N_m$, (b) transmitter-receiver distance $d = x_R - x_T$, (c) ion concentration $c_{ion}$ of the electrolyte medium, (d) average flow velocity $u$ inside the microfluidic channel.}
 \label{fig:SNR}
 \end{figure*}

\begin{figure*}[!t]
 \centering
   \subfigure[]{
 \includegraphics[width=4cm]{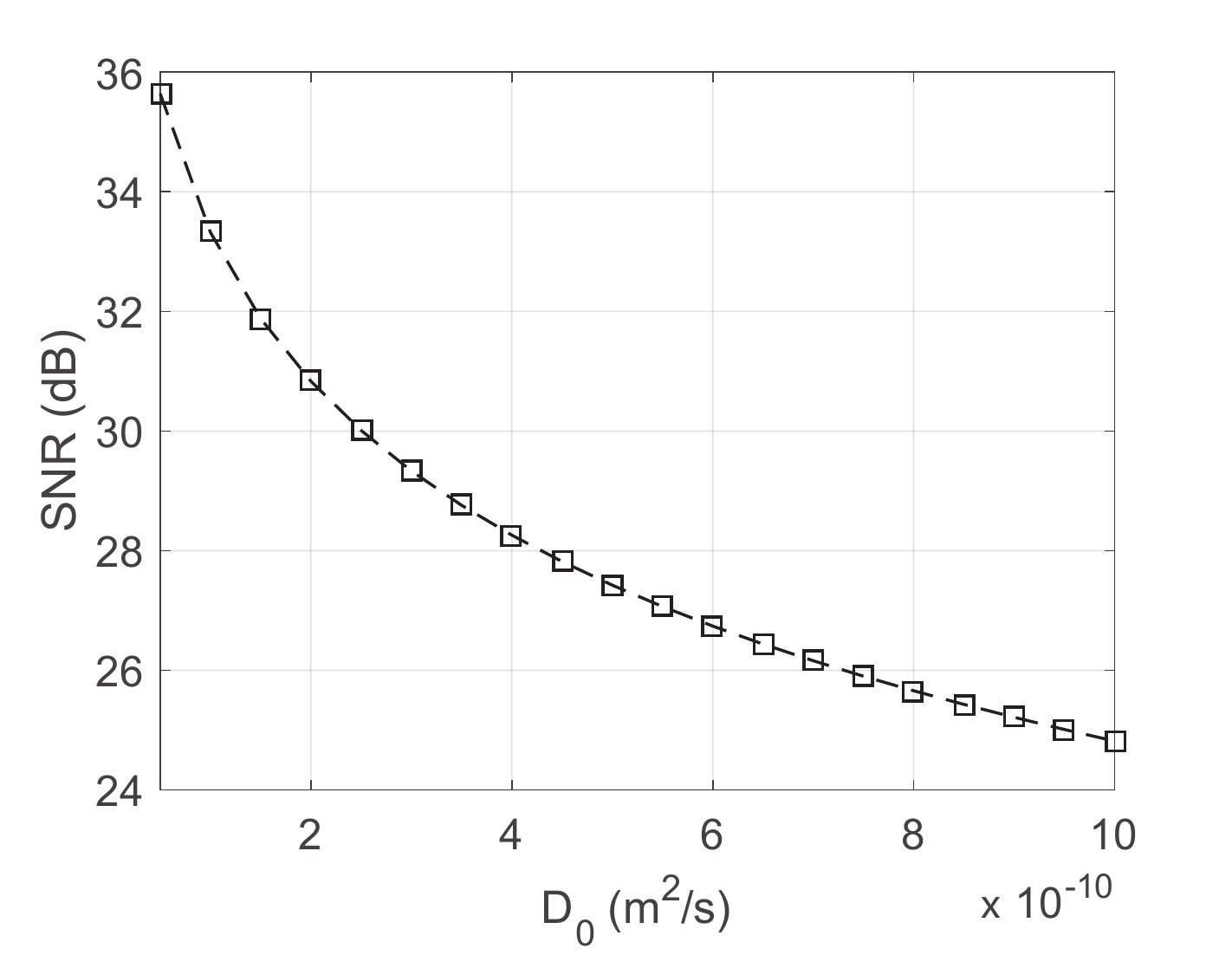}
 \label{fig:SNR_D0}
 }
  \subfigure[]{
 \includegraphics[width=4cm]{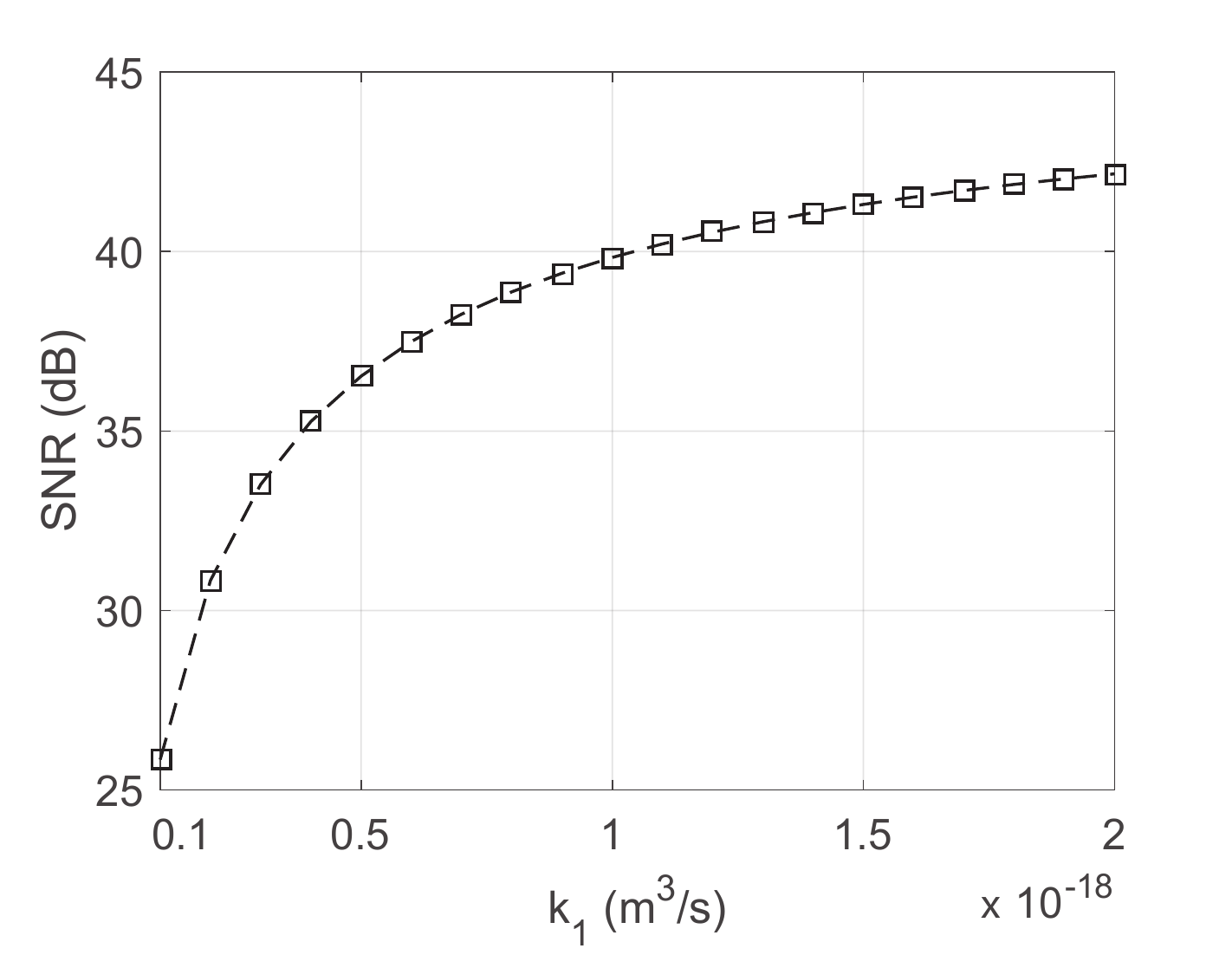}
 \label{fig:SNR_k1}
 }
    \subfigure[]{
 \includegraphics[width=4cm]{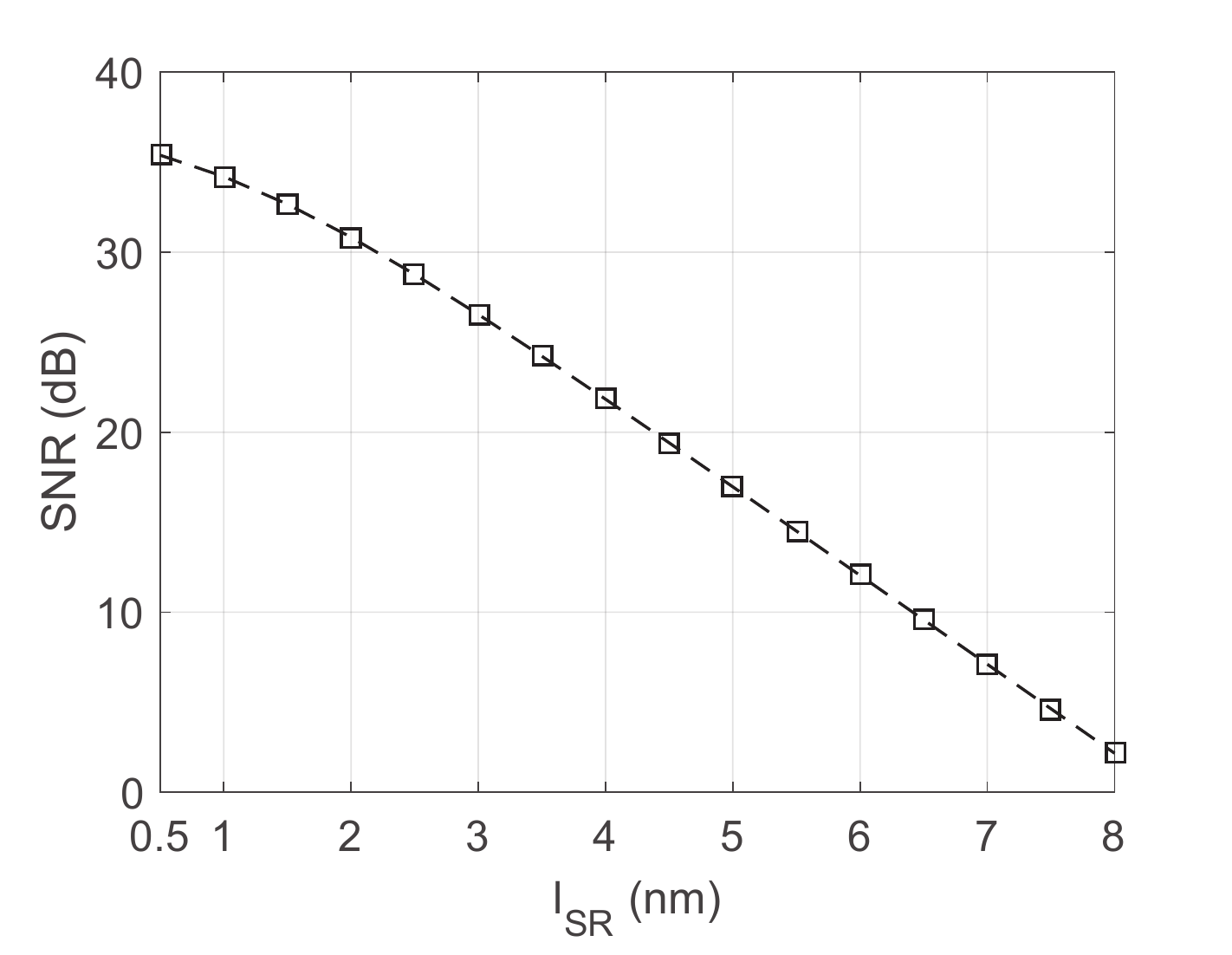}
 \label{fig:SNR_lsr}
 }
  \subfigure[]{
 \includegraphics[width=4cm]{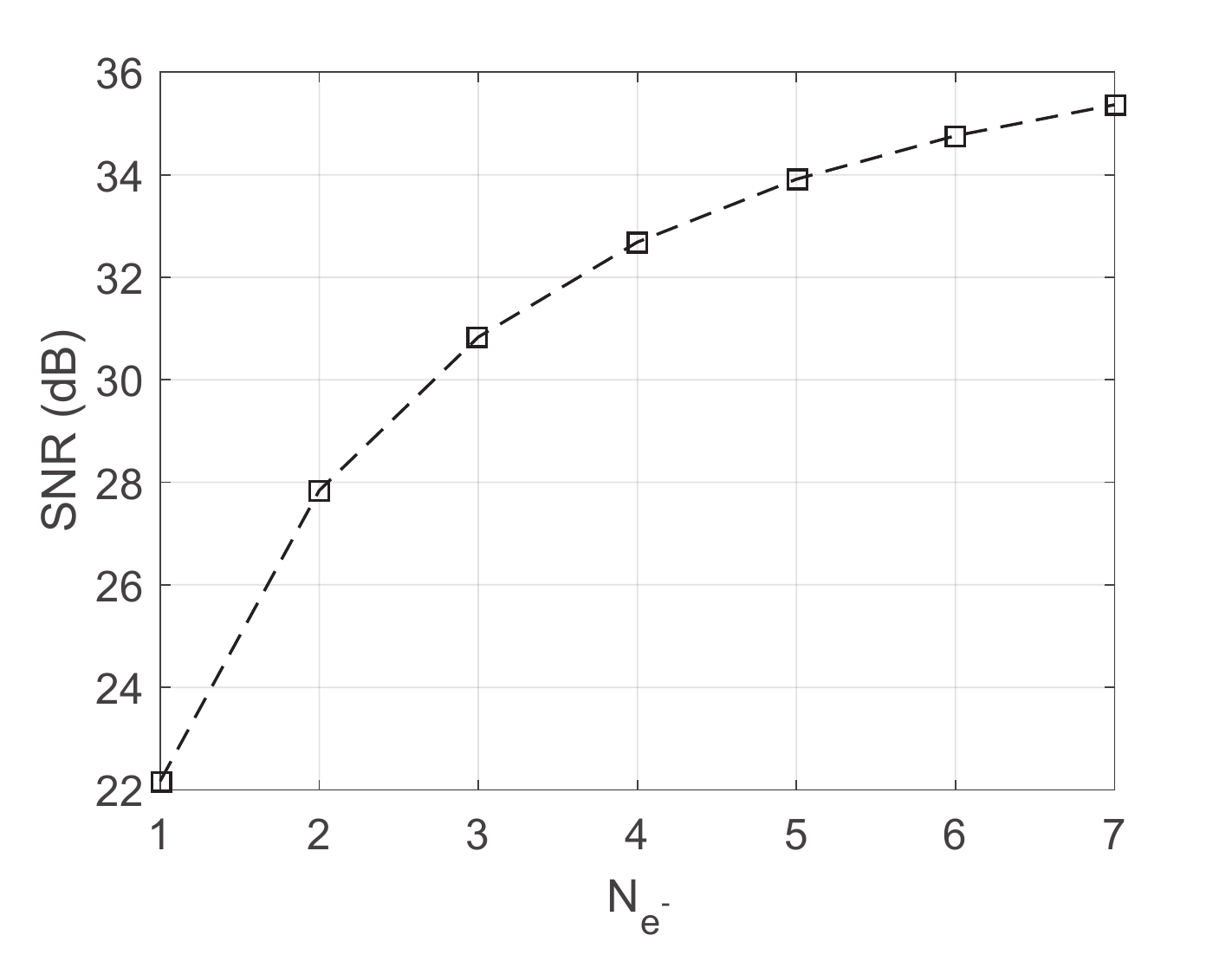}
 \label{fig:SNR_Ne}
 }
 \caption{Effect of the molecular parameters on the SNR at the electrical output of the receiver. SNR as a function of (a) intrinsic diffusion coefficient of ligands $D_0$, (b) intrinsic binding rate of ligands $k_1$, (c) surface receptor length $l_{SR}$, (d) number of free electrons per ligand molecule $N_e^-$.}
 \label{fig:SNR}
 \end{figure*}

\begin{figure*}[!t]
 \centering
   \subfigure[]{
 \includegraphics[width=4cm]{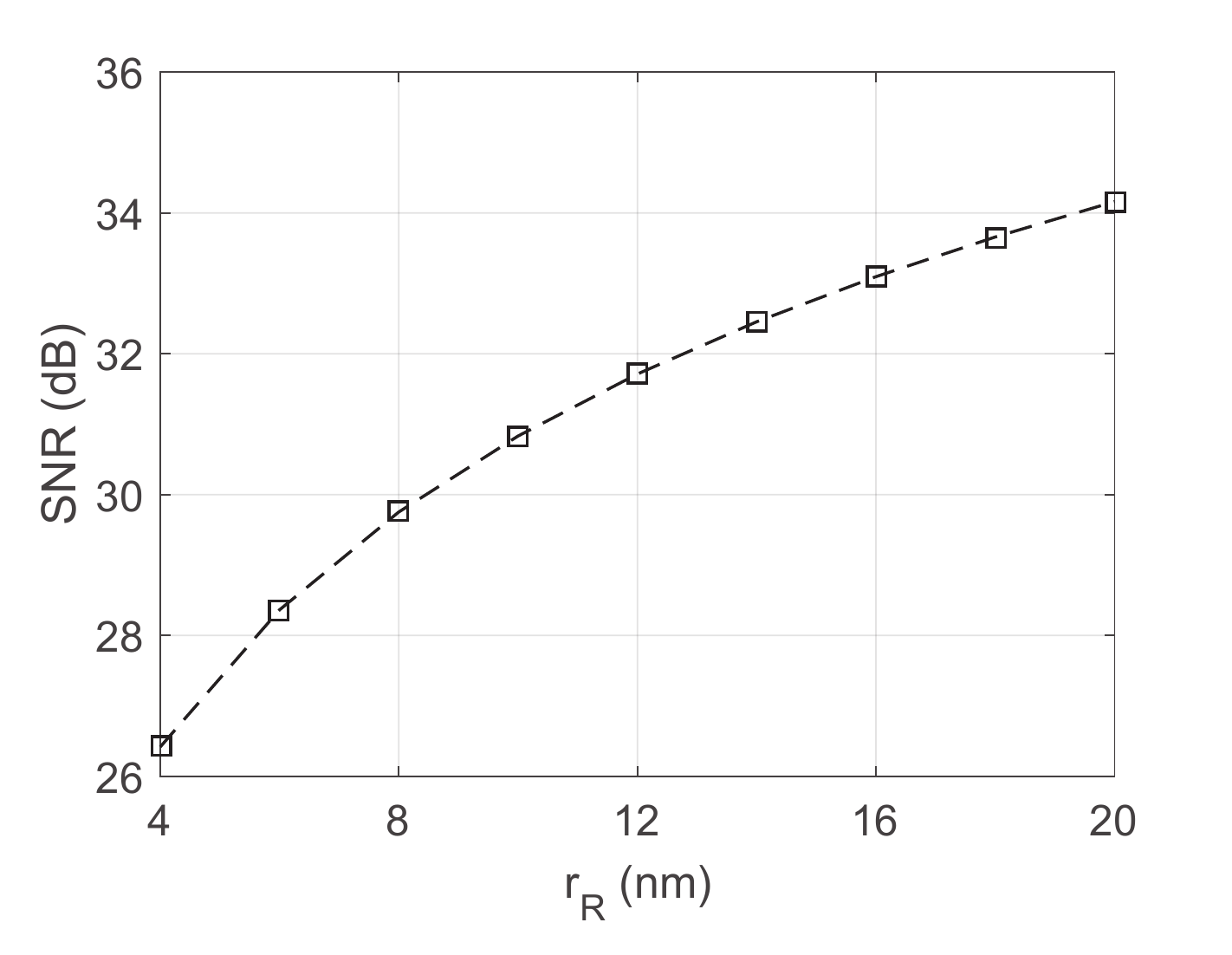}
 \label{fig:SNR_rR}
 }
  \subfigure[]{
 \includegraphics[width=4cm]{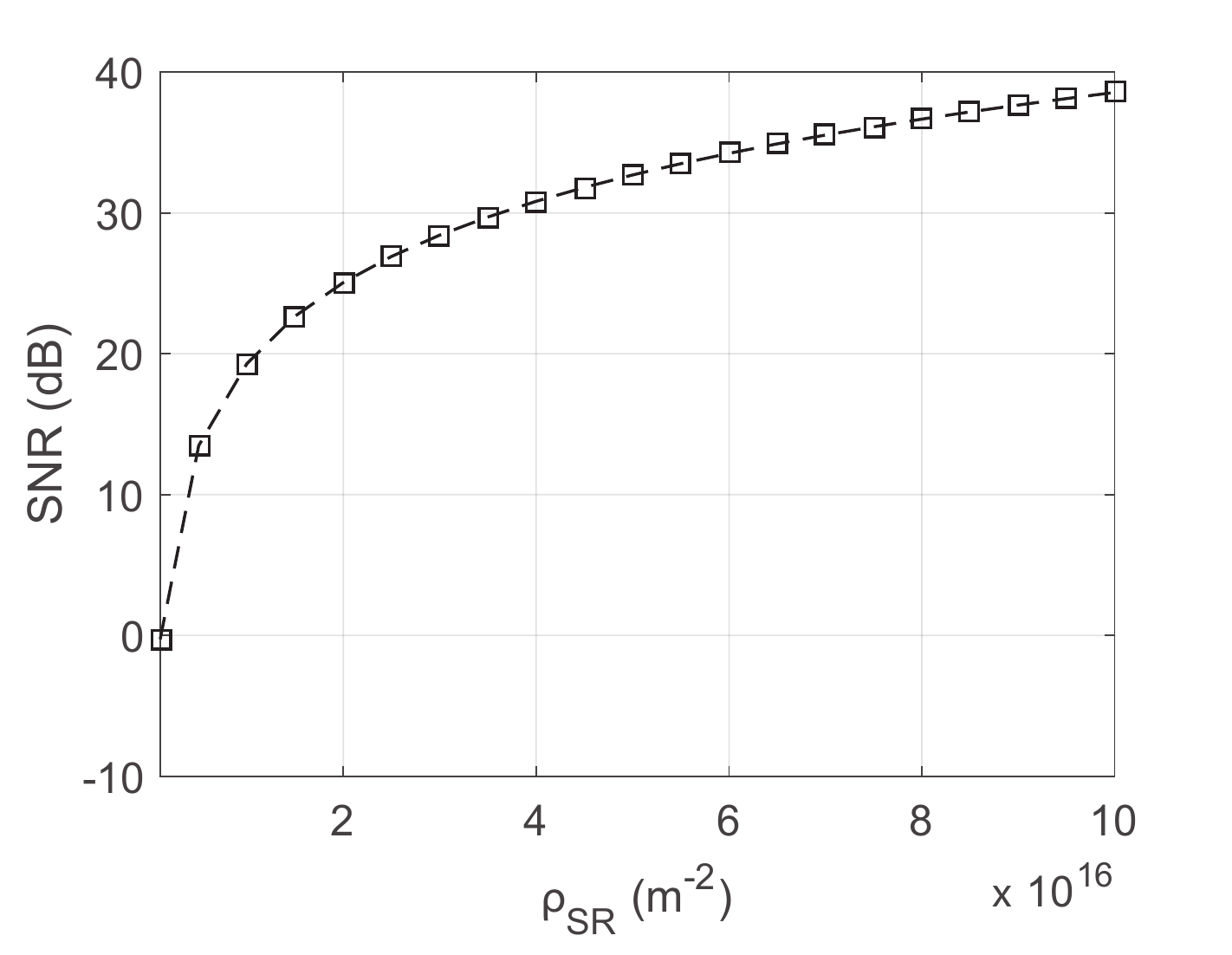}
 \label{fig:SNR_Nr}
 }
    \subfigure[]{
 \includegraphics[width=4cm]{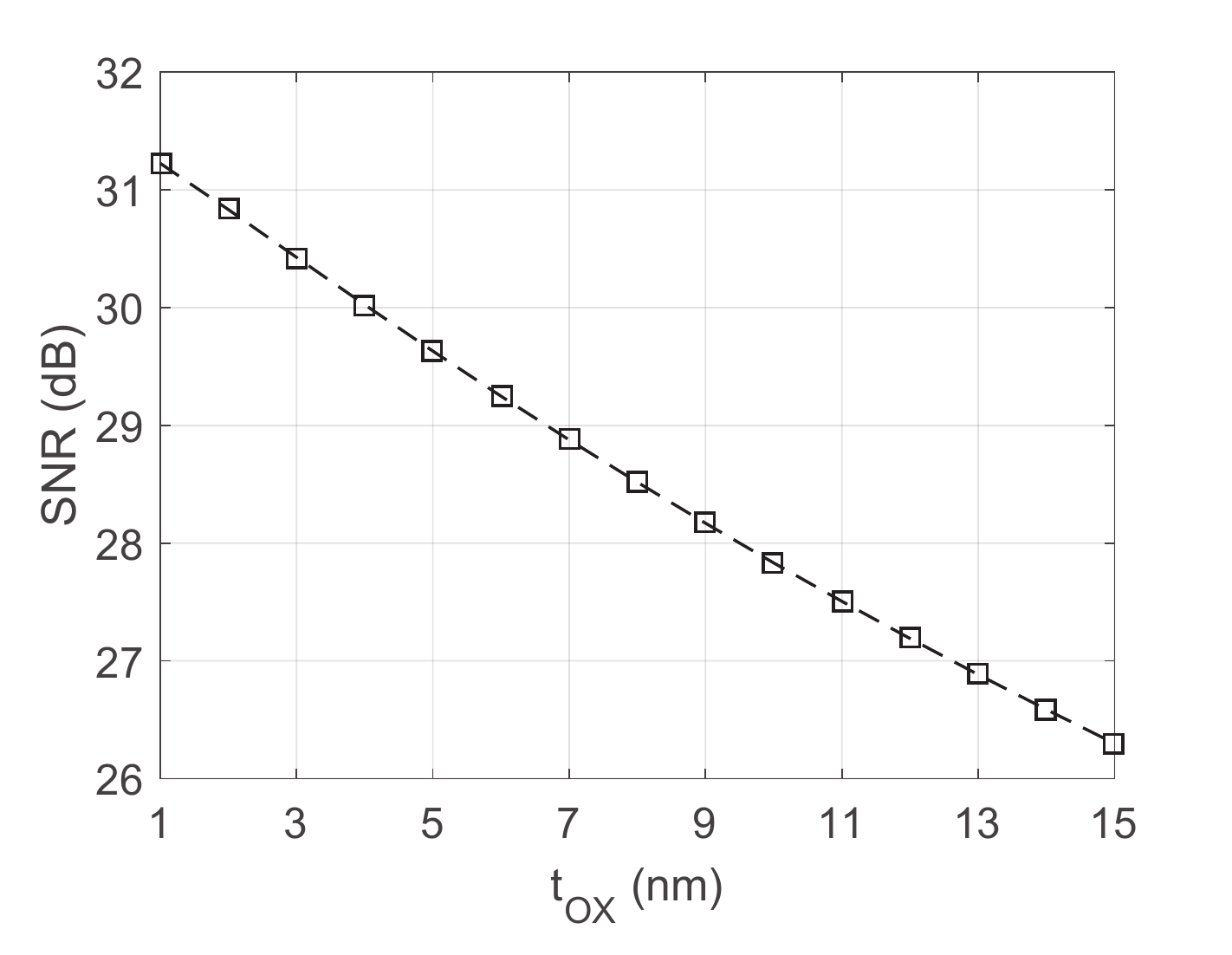}
 \label{fig:SNR_tox}
 }
  \subfigure[]{
 \includegraphics[width=4cm]{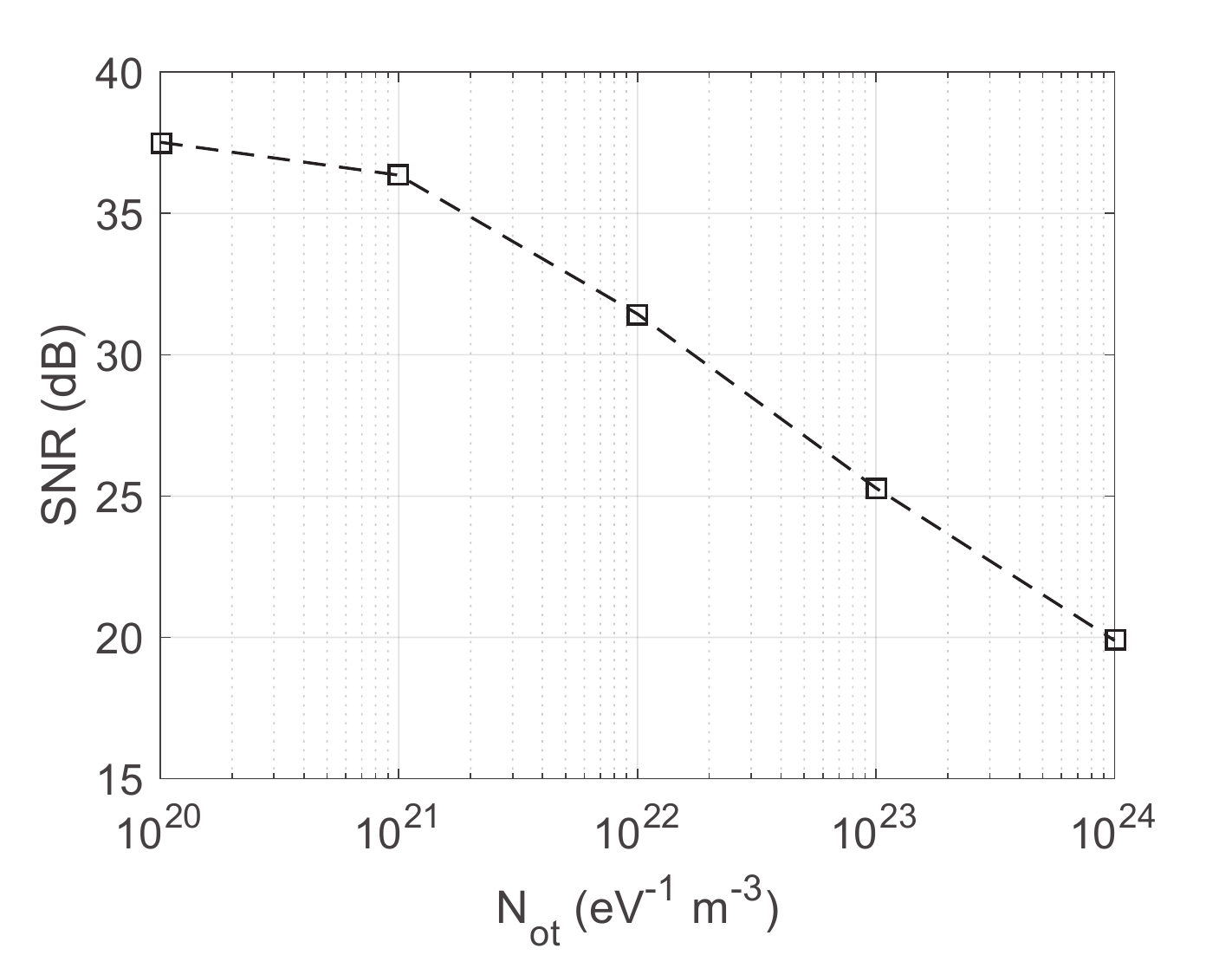}
 \label{fig:SNR_Not}
 }
 \caption{Effect of the receiver design parameters on the SNR at the electrical output of the receiver. SNR as a function of (a) SiNW radius $r_R$, (b) surface receptor concentration $\rho_{SR}$ (c) SiO$_2$ layer thickness $t_{OX}$, and (d) oxide trap density $N_t$ in SiNW.}
 \label{fig:SNR}
 \end{figure*}

\subsection{SNR Analysis}
In this section, we investigate the effect of main system parameters on the receiver's output SNR, which is formulated in \eqref{eq:SNRI}. We group the system parameters under three main categories: (i) communication system parameters related to the transmitter and communication channel, (ii) molecular parameters related to the characteristics of information carriers and corresponding receptors, (iii) receiver parameters related to the design of the SiNW FET-based MC receiver.

\subsubsection{Effect of Communication System Parameters}
SNR of the output current for varying number of ligands released by the transmitter is plotted in Fig. \ref{fig:SNR_Nm}, which clearly shows that SNR is significantly improved with increasing number of ligands. However, it begins to saturate at around $40$ dB for the default setting due to the saturation of the surface receptors for very high concentrations of ligands. Given a number of ligands released by the transmitter, the output SNR decreases with increasing transmitter-receiver distance $d$, as demonstrated in Fig. \ref{fig:SNR_d}. This is because the ligand concentration is attenuated (proportional to $\sqrt{d}$) as the distance increases.

The effect of ionic strength of the fluidic medium on the receiver SNR is demonstrated in Fig. \ref{fig:SNR_cion}. When the ionic concentration increases above 100 mol/m$^3$, the Debye length decreases below 1 nm resulting in substantial screening of ligand charges. Therefore, SNR significantly decreases with increasing ionic strength. Physiological conditions generally imply ionic concentrations higher than 100 mol/m$^3$. To compensate the attenuation of SNR, receptors with lengths comparable to Debye length should be selected.

The velocity of the fluid flow $u$ also has a remarkable effect on the SNR, as is seen Fig. \ref{fig:SNR_u}. As it increases, the ligands arrive more rapidly at the receiver, resulting in less attenuation. Therefore, higher velocity means higher ligand concentration at the receiver side, and this implies continuously improved SNR until it leads to the receptor saturation. As the velocity gets higher values, the transport rate of ligands $k_T$ to the receiver surface decreases because the flowing ligands could not find enough time to diffuse into the vicinity of the receptors. This in turn should decrease the SNR. Although this effect is well-captured by $k_T$, since we restrict the range of values that flow velocity can take for the sake of equilibrium assumption in this analysis, we could not observe the degrading effect of increasing flow velocity on the SNR.

\subsubsection{Effect of Molecular Parameters}
Diffusion coefficient $D_0$ is an important characteristic of the information ligands although it also depends on the temperature and the viscosity of the fluid. Its effect on the SNR is shown in Fig. \ref{fig:SNR_D0}. As is seen, the SNR decreases with increasing $D_0$. This is mainly caused by the increased dispersion of the ligands, which decreases the average ligand concentration that the receiver observes. The effect of binding constant, $k_1$, is plotted in Fig. \ref{fig:SNR_k1}. Increasing $k_1$ means that more ligands can bind to receptors as they flow over the receiver, and this results in an improved SNR. We also investigate the effect of receptor length, $l_{SR}$, when the ionic strength is set to 30 mol/m$^3$ which makes the Debye length equal to 1.75 nm. As seen in Fig. \ref{fig:SNR_lsr}, SNR in dB decreases almost linearly with the increasing length. The number of free charges per ligand also critically affects the receiver response, since the operation the SiNW transducer is mainly based on the field effect generated by the ligand charges. As demonstrated in \ref{fig:SNR_Ne}, employing highly charged ligands would improve the receiver SNR.
\begin{figure*}[!t]
 \centering
 \subfigure[]{
 \includegraphics[width=5.5cm]{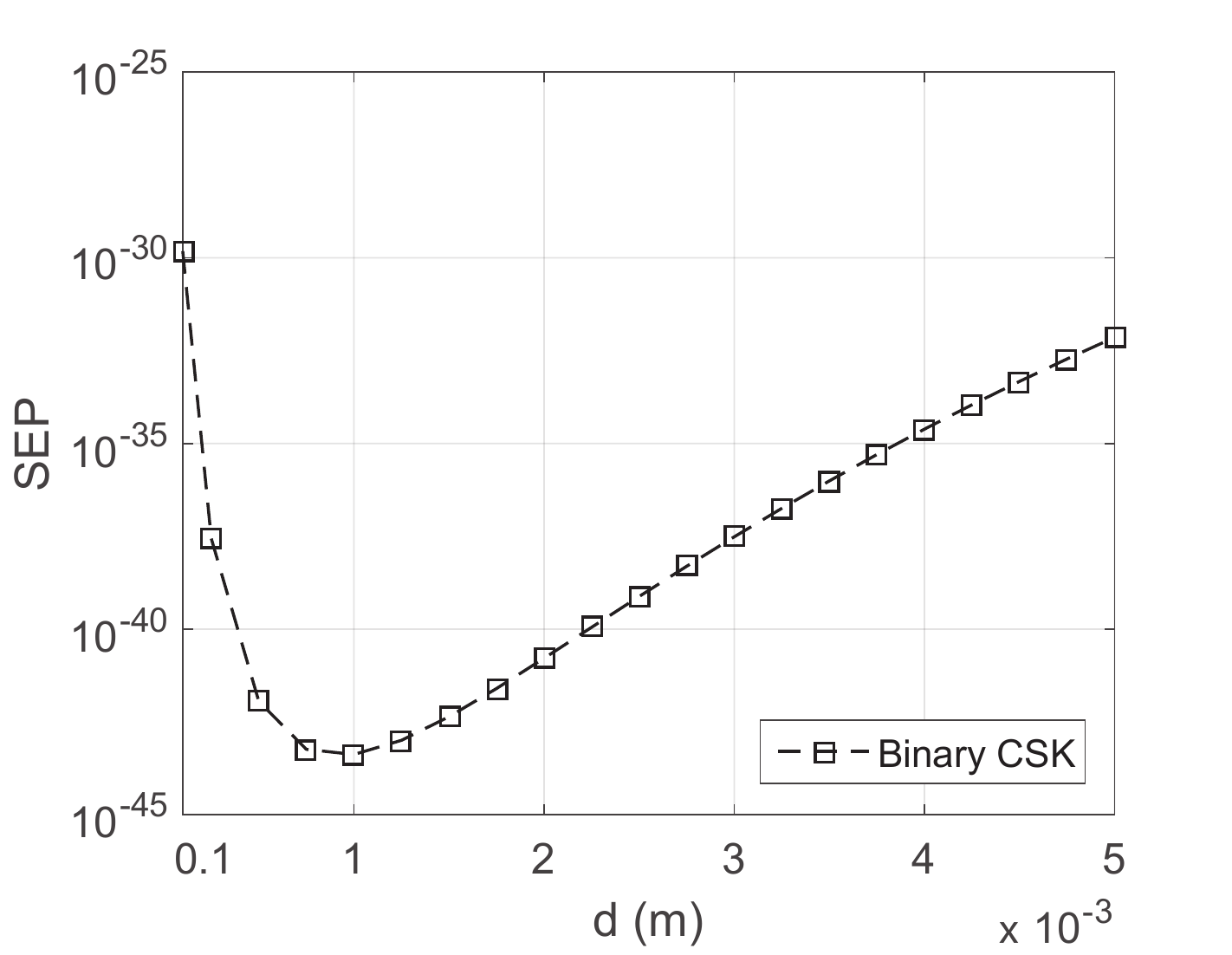}
 \label{fig:SEP_dB}
 }
 \subfigure[]{
 \includegraphics[width=5.5cm]{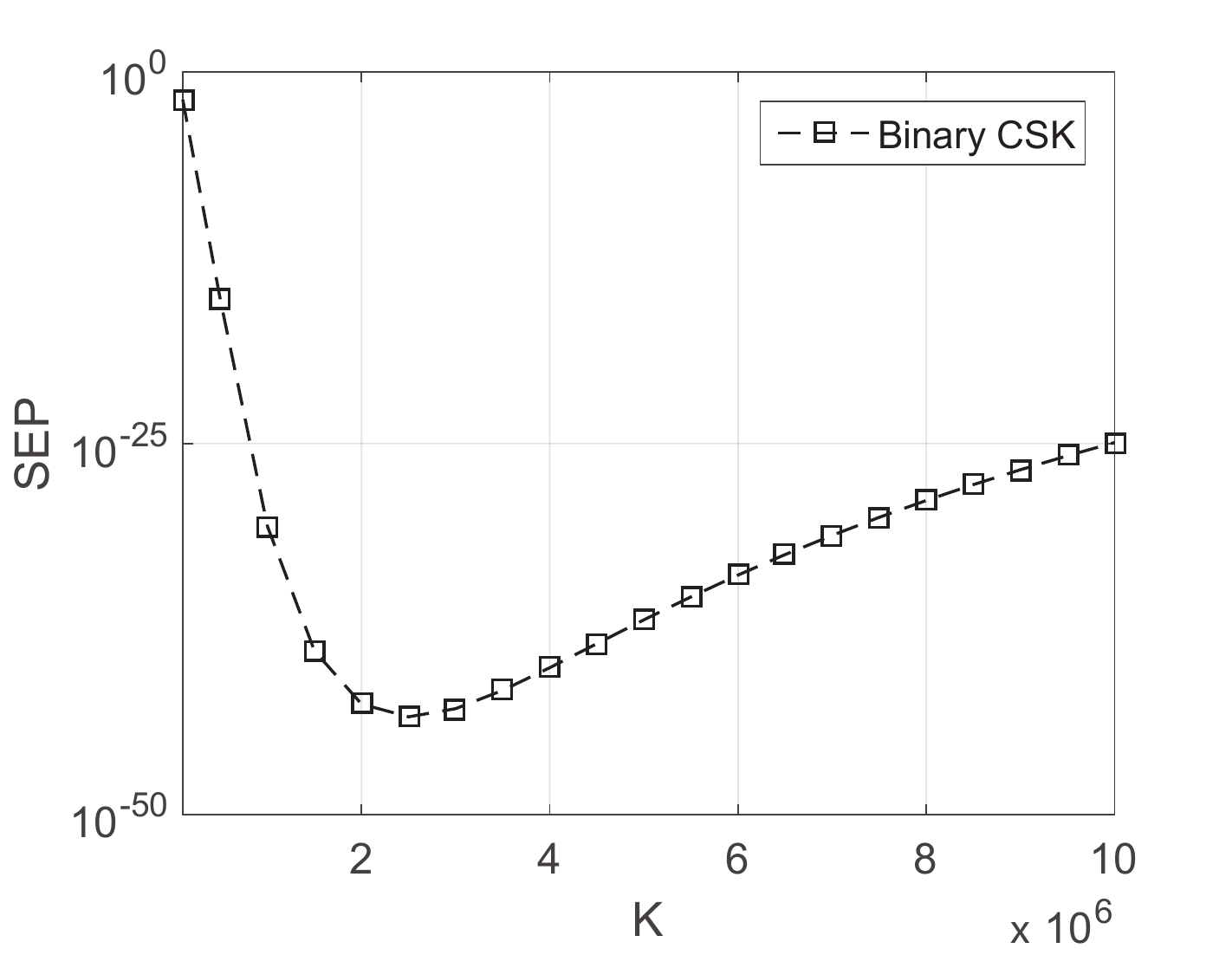}
 \label{fig:SEP_KB}
 }
 \subfigure[]{
 \includegraphics[width=5.5cm]{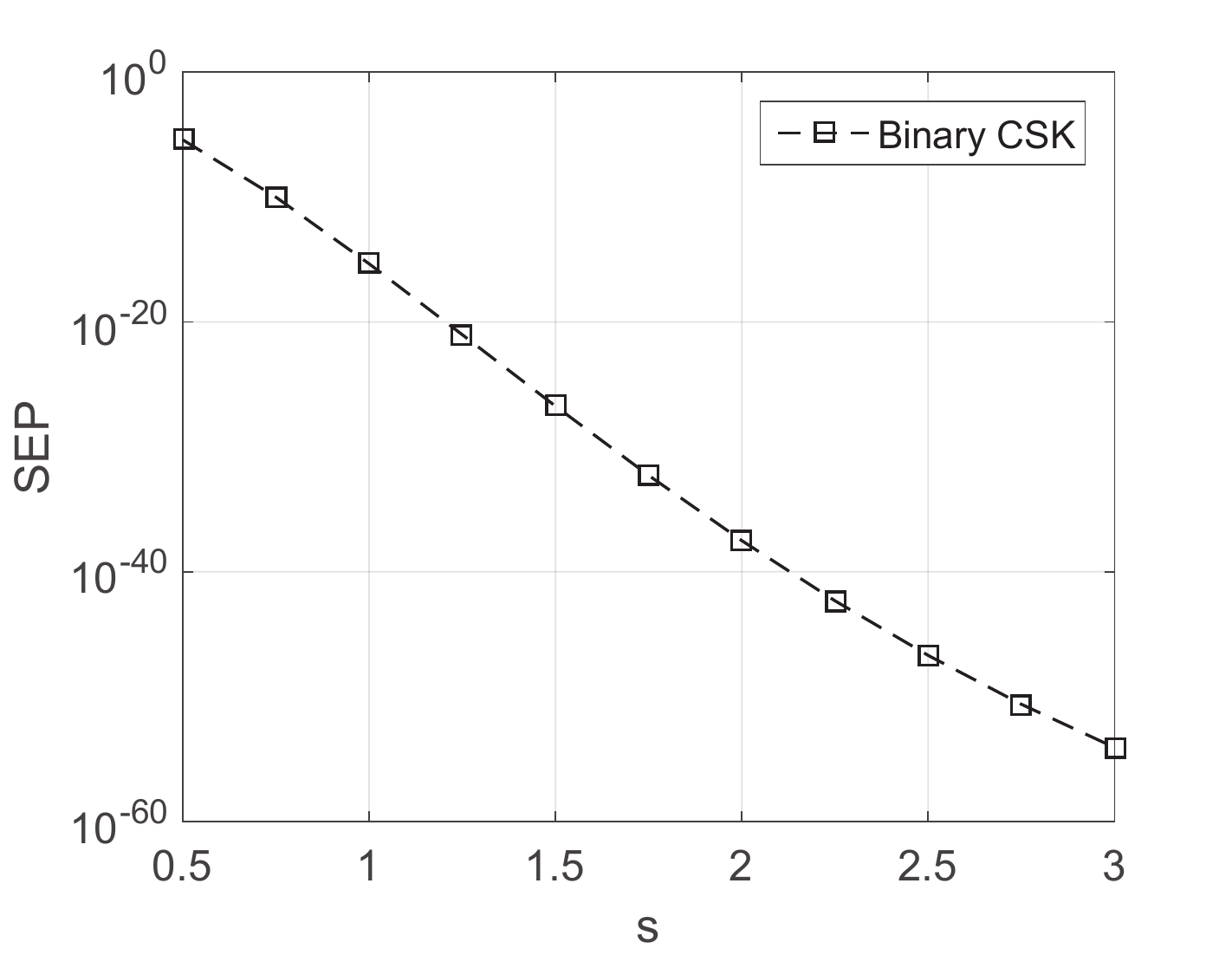}
 \label{fig:SEP_sB}
 }
 \subfigure[]{
 \includegraphics[width=5.5cm]{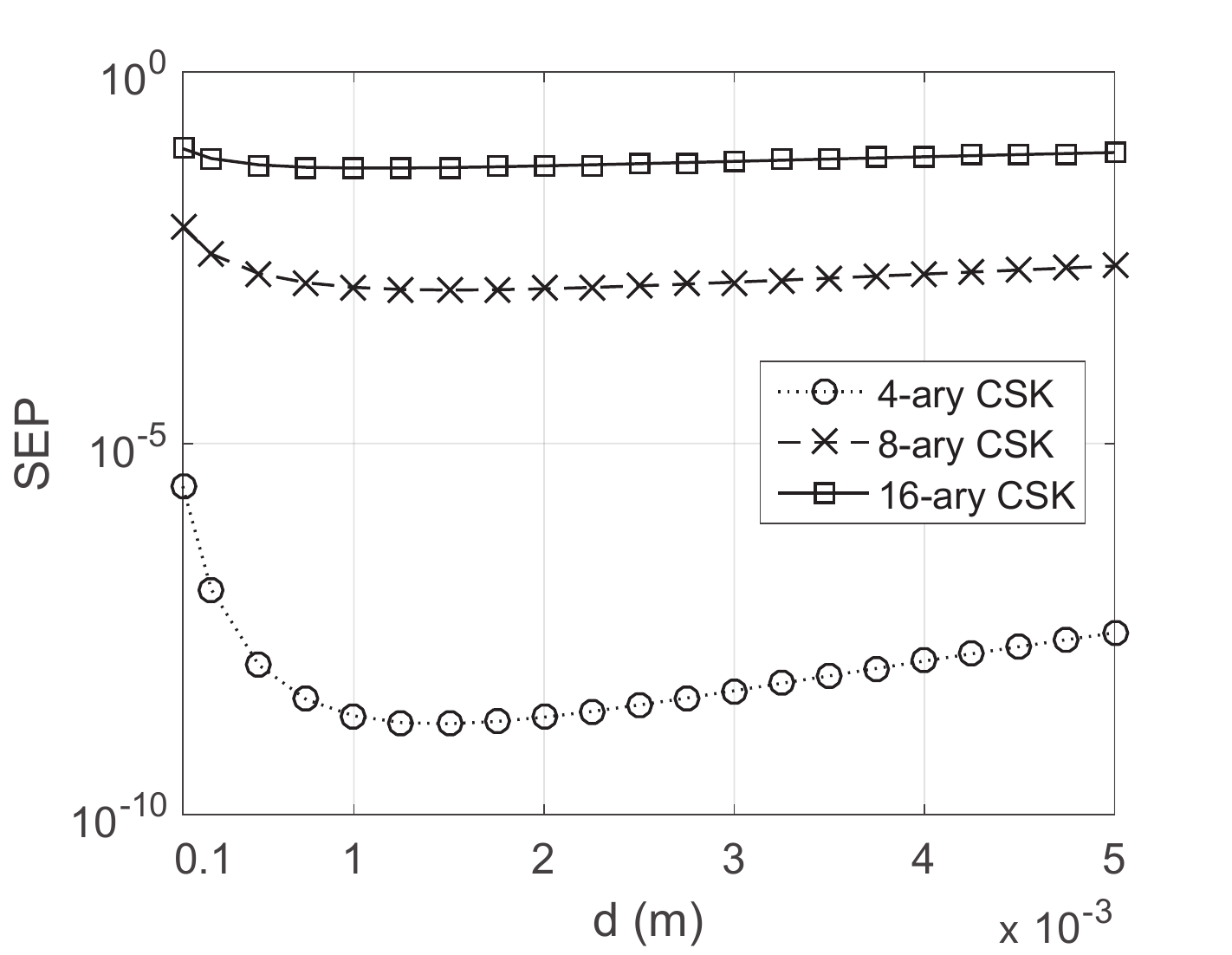}
 \label{fig:SEP_dM}
 }
 \subfigure[]{
 \includegraphics[width=5.5cm]{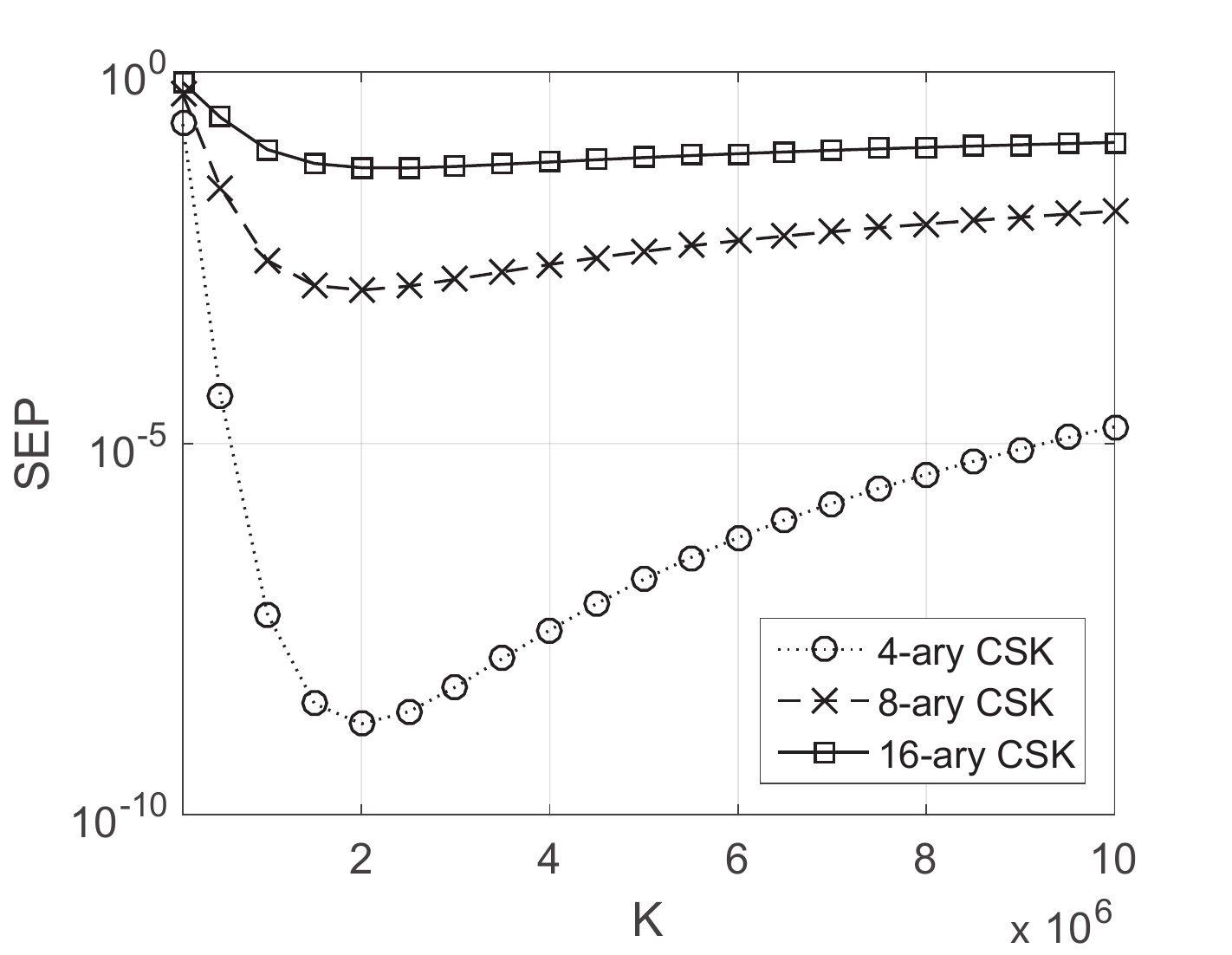}
 \label{fig:SEP_KM}
 }
 \subfigure[]{
 \includegraphics[width=5.5cm]{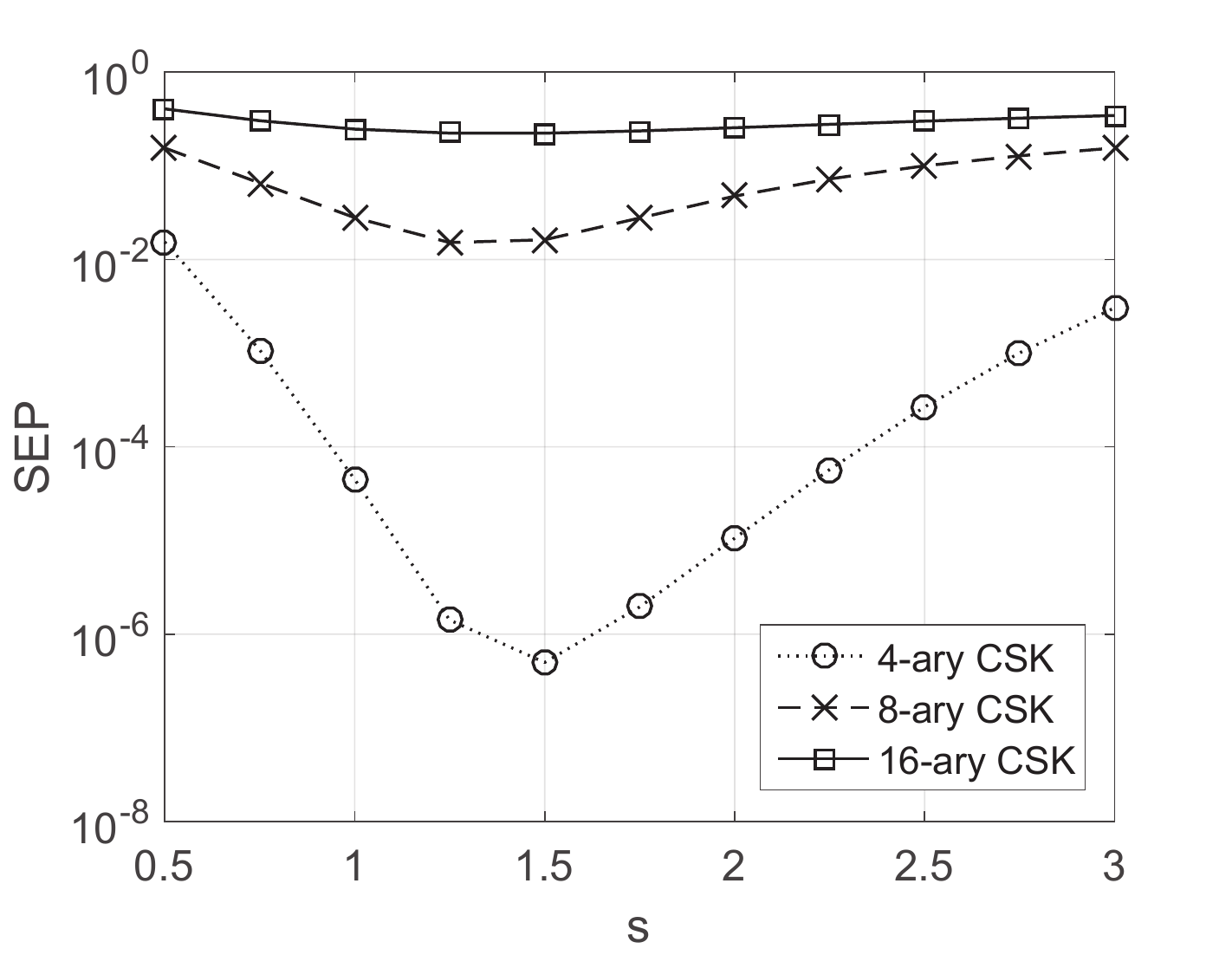}
 \label{fig:SEP_sM}
 }
 \caption{Symbol error probability (SEP) for M-CSK modulation. SEP as a function of (a, c) transmitter-receiver distance $d$, (b, e) maximum number $K$ of ligands that transmitter can release, (c, f) constellation exponent $s$.}
 \label{fig:SEN}
 \end{figure*}

\subsubsection{Effect of Receiver Parameters}
An important parameter of the receiver design is the size of the SiNW, which has a direct impact on the the ligand flux to the receiver, the number of surface receptors, and the capacitance values of the oxide layer, the SiNW double layer and the diffusion layer. Since we fixed the length of the SiNW as equal to the cross-sectional length of the microfluidic channel, we only change its radius $r_R$, and thus, its effective width $w_R$ (recall that $w_R \simeq \pi \times r_R$). The effect of SiNW radius is demonstrated in Fig. \ref{fig:SNR_rR}. Increasing the radius also increases the SNR. This is mainly because the transport rate of the ligands $k_T$ and the number of surface receptors $N_R$ increase with the radius.

Given that the SiNW radius is fixed to its default value $r_R = 10$ nm, increasing the receptor concentration $\rho_{SR}$, i.e., increasing the number of surface receptors (note that $N_R = w_r \times \rho_{SR}$), is another way of improving SNR, as can be inferred from Fig. \ref{fig:SNR_Nr}. However, restrictions imposed by the receptor size and possible interactions among densely deployed receptors, such as negative cooperativity, which are not captured by this model, should be accounted for in a real world implementation. We also analyze the effect of oxide layer thickness $t_{OX}$, which determines the oxide layer capacitance $C_{OX} = \epsilon_{OX}/t_{OX}$. Fig. \ref{fig:SNR_tox} demonstrates the results for conventional values of $t_{OX}$. As can be inferred, lower $t_{OX}$ implies an improved SNR for the default system settings used in our analysis. This is mainly because increasing $C_{OX}$ results in higher values of transconductance $g_{FET}$, which means more effective transduction of the surface potential to the output current. On the other hand, the effect of $C_{OX}$ on the equivalent capacitance of the transducer is usually negligible compared to the substantial effect of the diffusion layer capacitance $C_{DL}$. Lastly, we analyze the SNR for varying oxide trap density $N_{ot}$, which is proportional to the impurity of the SiNW. Trap density affects the carrier mobility, and increases the $1/f$ noise. Negative effect of increasing trap density on the SNR is evident from Fig. \ref{fig:SNR_Not}.

\subsection{SEP Analysis}
In the last analysis, we evaluate the performance of the receiver when M-CSK is utilized for the modulation. We find the SEP for binary, 4-ary, 8-ary, 16-ary cases for different system settings. For better visualization, we present the results for each setting in two different plots separating the binary case from the 4-, 8-, 16-ary cases. For the constellation design, we assume
\begin{equation}
N_m = \ceil*{(m+1)^s \times \left(K/M^s\right)}, \label{modulation}
\end{equation}
where $K$ is the maximum number of molecules that TN can release in a single transmission, $s$ is the exponent defined to obtain a non-uniform constellation, and $m\in\mathcal{M} = \{0, 1,...,M-1\}$ with $M \in \{2, 4, 8, 16\} $. Except for the analysis where we investigate the effect of distance, we set $K = 1 \times 10^6$ and $s=1$, so that we obtain a uniform constellation where the adjacent symbols are separated by $K/M$ number of ligands. For the distance analysis, to be able to reveal the correlated effect of the distance and maximum number of ligands, we set $K = 4 \times 10^6$.
\begin{figure*}[!t]
 \centering
 \subfigure[]{
 \includegraphics[width=5.5cm]{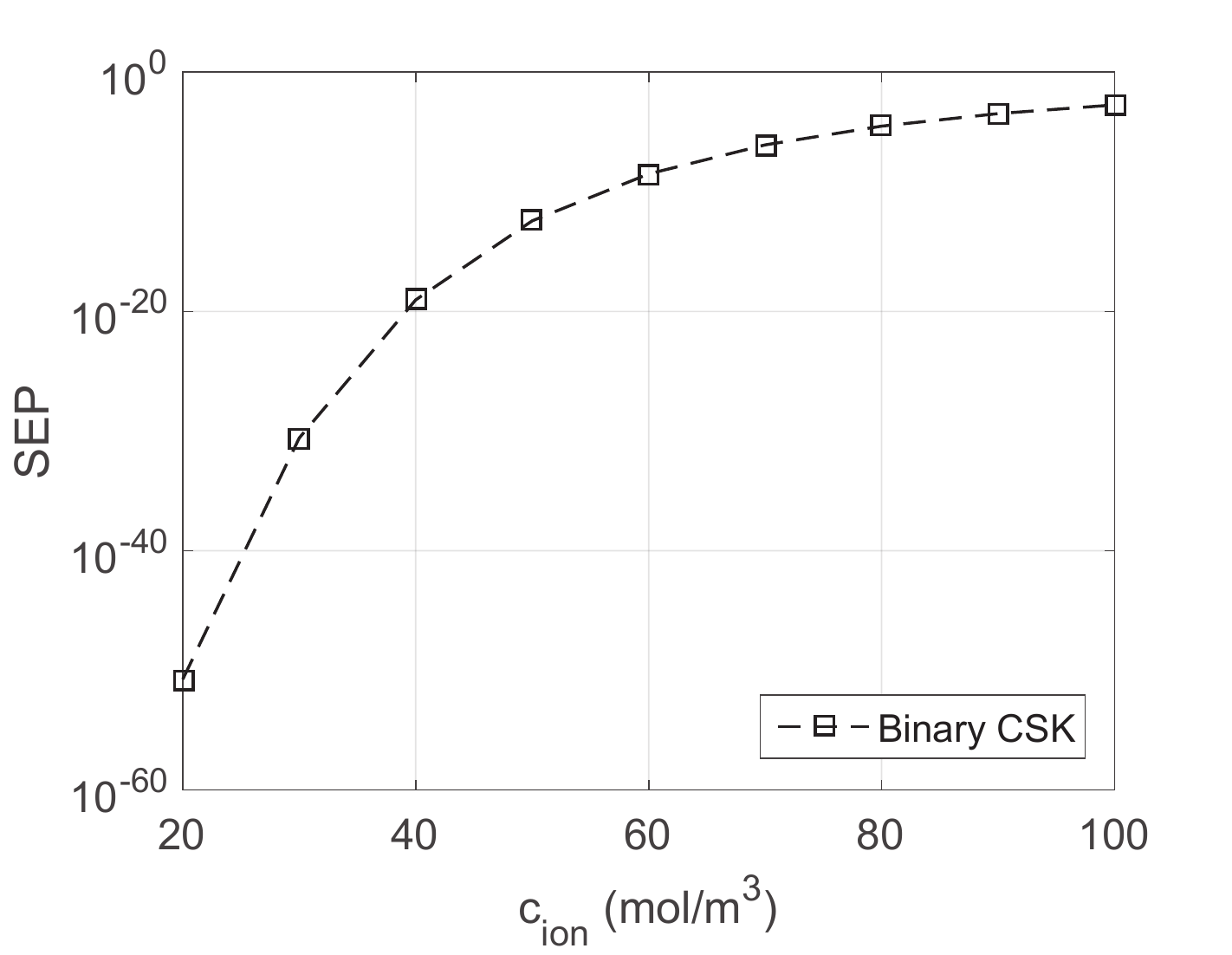}
 \label{fig:SEP_cionB}
 }
 \subfigure[]{
 \includegraphics[width=5.5cm]{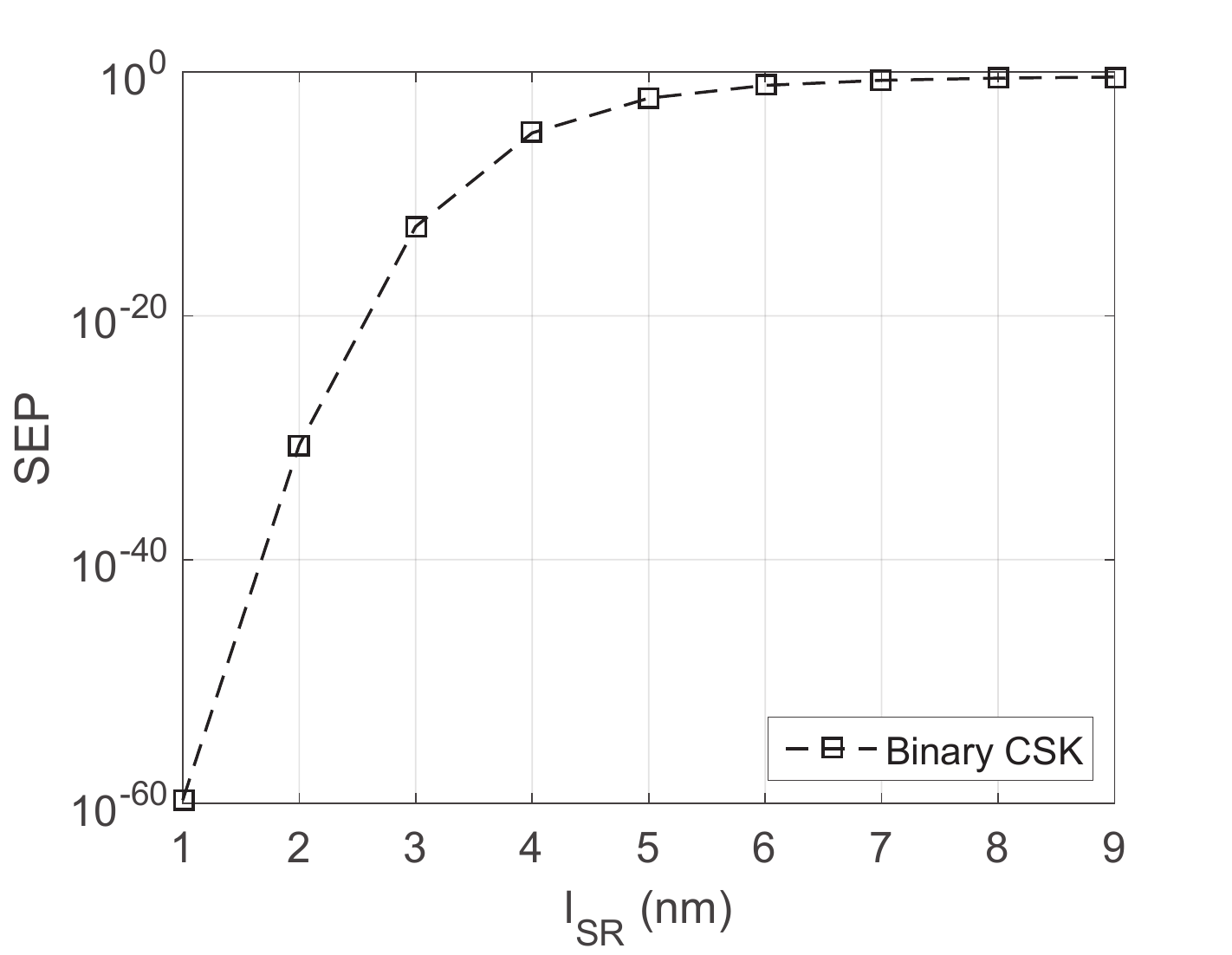}
 \label{fig:SEP_lrB}
 }
 \subfigure[]{
 \includegraphics[width=5.5cm]{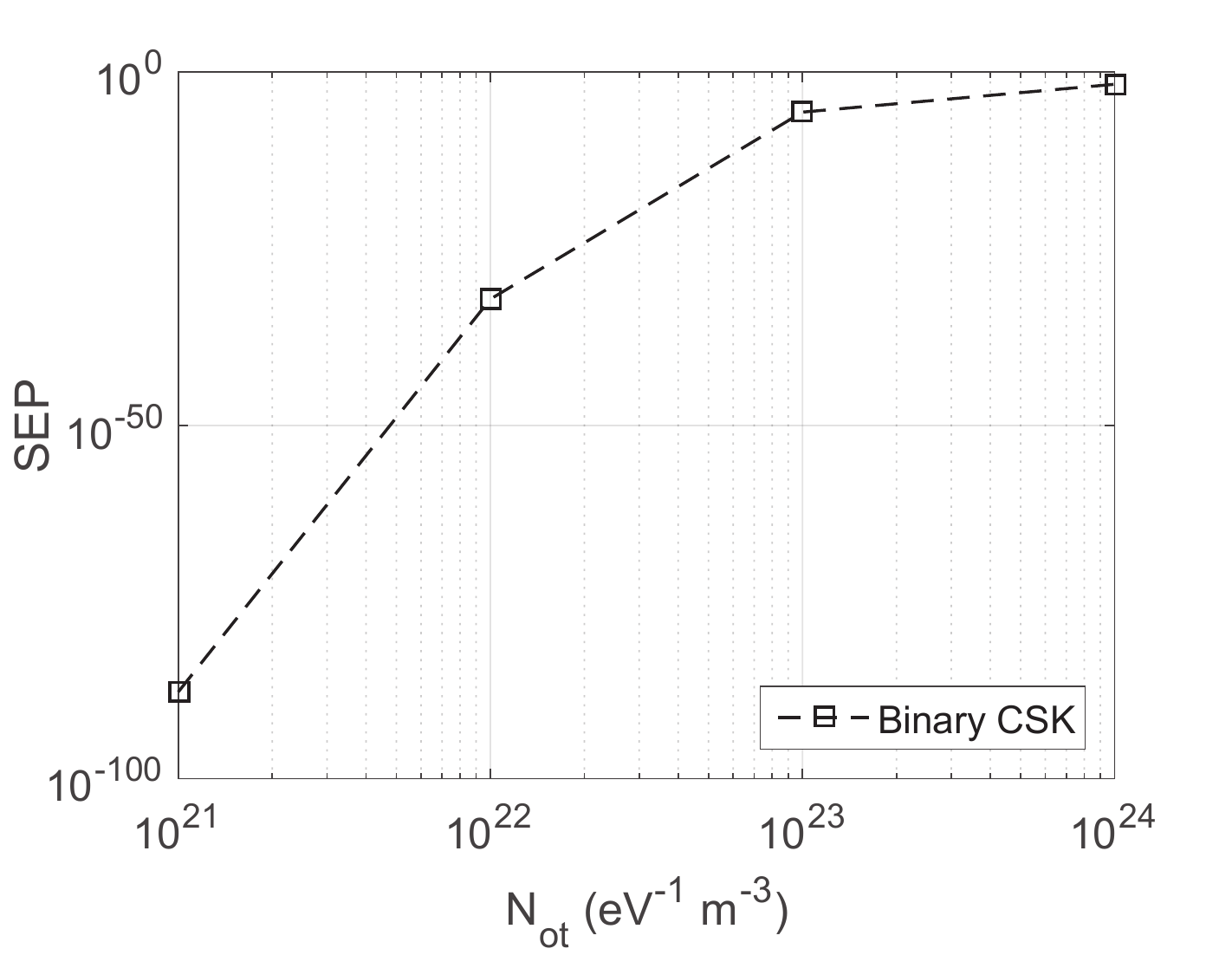}
 \label{fig:SEP_NtB}
 }
 \subfigure[]{
 \includegraphics[width=5.5cm]{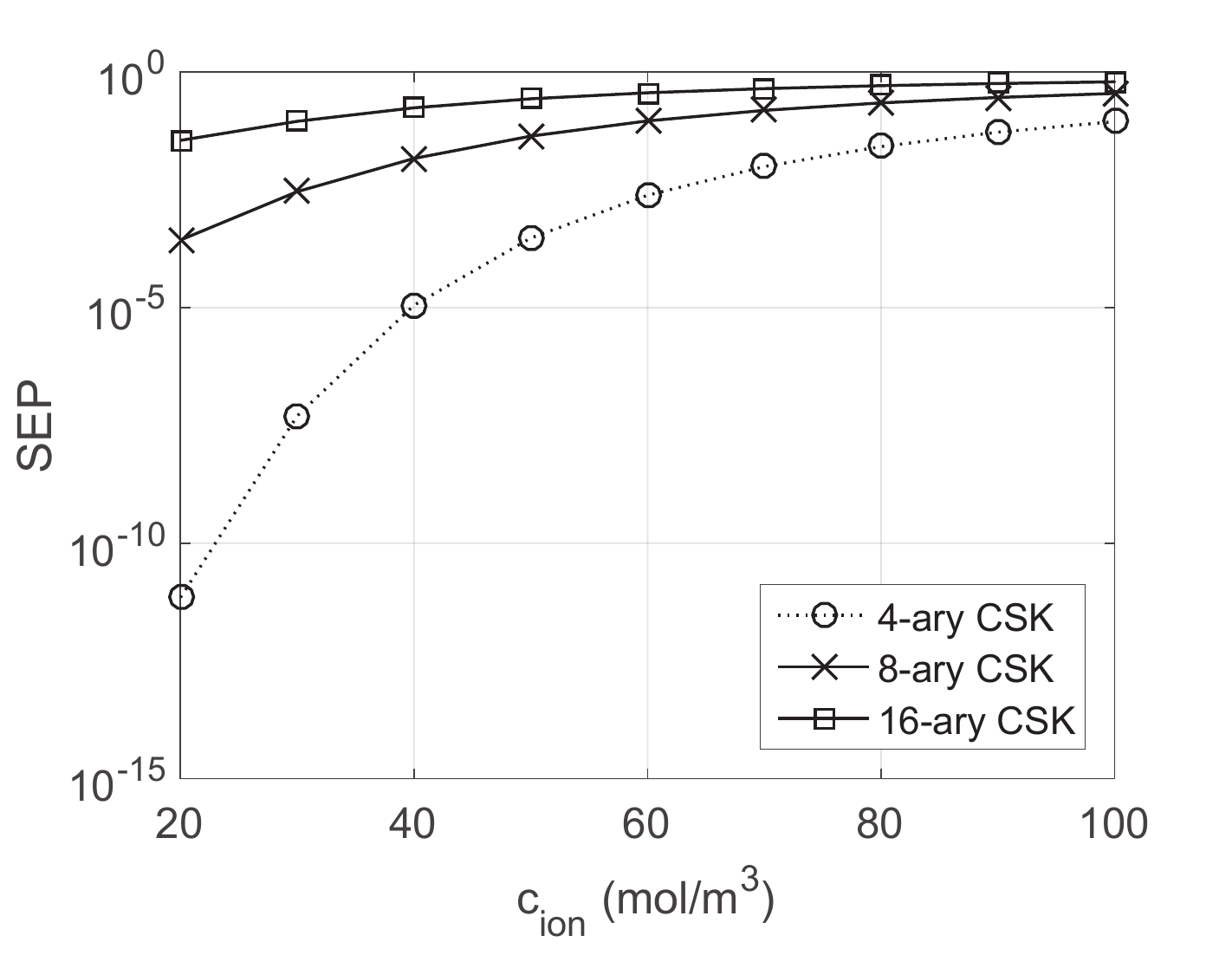}
 \label{fig:SEP_cionM}
 }
 \subfigure[]{
 \includegraphics[width=5.5cm]{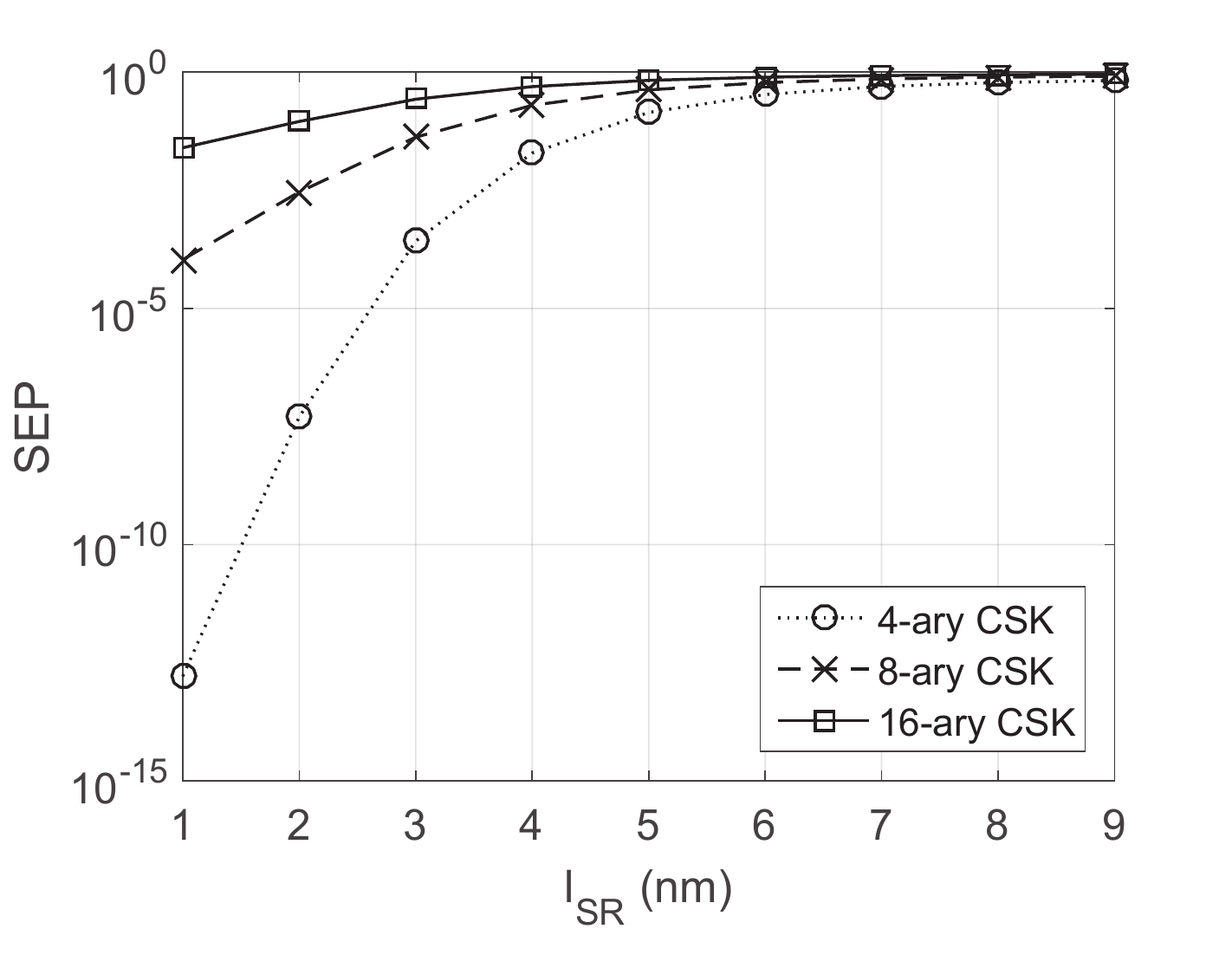}
 \label{fig:SEP_lrM}
 }
 \subfigure[]{
 \includegraphics[width=5.5cm]{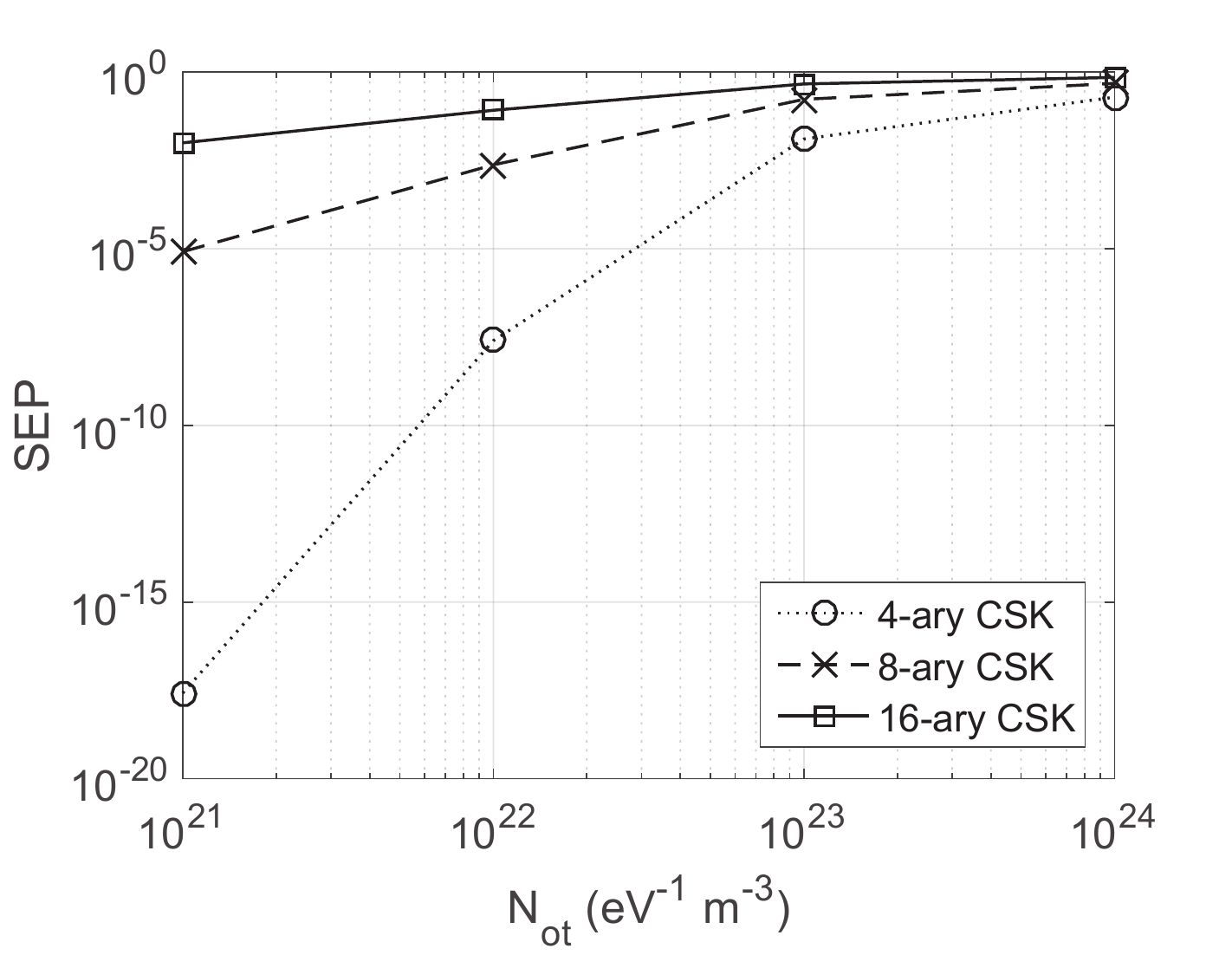}
 \label{fig:SEP_NtM}
 }
 \caption{Symbol error probability (SEP) for M-CSK modulation. SEP as a function of (a, c) ion concentration $c_{ion}$ of the electrolyte medium, (b, e) receptor length $l_{SR}$, (c, f) oxide trap density $N_{ot}$ in SiNW.}
 \label{fig:SEN}
 \end{figure*}

Figs. \ref{fig:SEP_dB} and \ref{fig:SEP_dM} show the SEP as a function of transmitter-receiver distance $d$. As is seen, for all modulation schemes, the SEP is minimum for intermediate distances, e.g., $1-2$ mm, and begins to increase when the distance is below or above this range. The reason can be explained as follows. As the distance gets smaller, the receiver operates near saturation because the concentration of ligands at the receiver location significantly increases. This is reflected to the output current, and results in a reduced sensitivity such that the receiver cannot discriminate different levels of ligand concentration corresponding to different symbols. In a similar way, when the distance is increased, ligand concentration is substantially attenuated until the molecular signal reaches the receiver, which also leads to a degradation in receiver sensitivity. Hence, we can conclude that there is an optimal range of distance for a given $K$.

Next, we analyze the effect of maximum number of ligands that the transmitter can release. As can be inferred from the results presented in Figs. \ref{fig:SEP_KB} and \ref{fig:SEP_KM}, increasing $K$ up to $2 \times 10^6$ decreases the SEP for binary CSK. However, when we further increase $K$ above this range, the SEP begins to get higher. Similar trends can be observed for 4-, 8-, and 16-ary cases. The reason is closely related to the reasons that lead to the results obtained with varying distance in the previous analysis. As the transmitter releases higher number of ligands for each symbol, the receiver begins to operate near saturation, which degrades its ability to discriminate different symbols. The nontrivial results obtained for varying distance and maximum number of transmitted ligands are the consequences of the dynamic range imposed by the receiver, which has a limited reception capacity set by the number of surface receptors. Similar trends were noted in MC experiments conducted with metal oxide semiconductor alcohol sensors \cite{Farsad2013}, \cite{Kim2015}. Although these sensors are not operating based on ligand-receptor binding mechanism, they result in saturation when exposed to high concentrations of alcohol, since the devices have an active channel limited in size.

Since the receiver response is nonlinear as obvious from Fig. \ref{fig:signal}, utilizing a uniform constellation for M-CSK modulation is not optimal. Particularly, there is a need to place more symbols on the lower half of the modulation range, for which the receiver is more sensitive to concentration variations. We analyze the performance employing a simple non-uniform M-ary modulation scheme by varying the exponent $s$ in \eqref{modulation}. For binary CSK, rising the exponent significantly decreases SEP, as demonstrated in Fig. \ref{fig:SEP_sB}. However, when more than two symbols are transmitted, different trends are observed in Fig. \ref{fig:SEP_sM}. After some threshold, the SEP begins to increase again. This is because the low level symbols approach to each other when we increase $s$, which complicates the detection. Hence, we need more complex constellation designs that can properly exploit the nonlinear response of the receiver.

Second set of analyses is performed for parameters related to the receiver and propagation medium. In Figs. \ref{fig:SEP_cionB} and \ref{fig:SEP_cionM}, we see that increasing the ionic concentration $c_{ion}$ substantially degrades the receiver performance for all M-CSK modulation schemes. Screening of the ligand charges by the medium counterions leads to a lower signal power at the receiver output. In the presence of signal-independent $1/f$ noise, this is reflected to a degraded detection performance of the receiver. The same reason leads to the results obtained for varying receptor lengths, which are presented in Figs. \ref{fig:SEP_lrB} and \ref{fig:SEP_lrM}. As can be inferred, it is possible to obtain a SEP lower than $10^{-10}$ for binary CSK by utilizing receptors with $l_{SR}<3$ nm. Trap density of SiNW channel influences the extent of the $1/f$ noise. Lower values of $N_{ot}$ indicate a clean semiconductor channel and lead to an improved receiver performance, as evident from Figs. \ref{fig:SEP_NtB} and \ref{fig:SEP_NtM}.

\section{Conclusion}

In this paper, we have developed a communication theoretical model for SiNW FET-based MC receivers. Focusing on microfluidic MC, we have derived closed-form expressions for fundamental performance metrics, such as SNR and SEP, to provide an analysis and optimization framework for MC with nanobioelectronic receivers. The results of performance evaluation pointed out several optimization pathways that need to be taken to improve the detection performance. The model can be extended to incorporate the transient dynamics of MC and bioFETs for enabling analysis also in the frequency domain. Open issues include the design of optimal constellations and detection schemes for MC systems equipped with bioFET receivers.

\end{document}